\documentclass[emulateapj,psfig,apjfonts]{emulateapj}
\usepackage{natbib} \usepackage{subfigure}
\usepackage{graphicx}

\begin{document}

\title{Tracers of Chromospheric Structure I: Observations of Ca II K and H$\alpha$
in M Dwarfs}

\author{Lucianne M. Walkowicz}  \affil{Astronomy Department,
University of Washington, Box 351580, Seattle, WA 98195-1580}
\email{lucianne@astro.washington.edu}

\author{Suzanne L. Hawley}  \affil{Astronomy Department, University of
Washington, Box 351580, Seattle, WA 98195-1580}
\email{slh@astro.washington.edu}

\begin{abstract}
We report on our observing program\footnote{This paper is based on
observations obtained  with the Apache Point Observatory 3.5-meter
telescope, which is owned and operated by the Astrophysical Research
Consortium.  Some of the data presented herein were obtained at the
W.M. Keck Observatory, which is operated as a scientific partnership
among the California Institute of Technology, the University of
California and the National Aeronautics and Space Administration. The
Observatory was made possible by the generous financial support of the
W.M. Keck Foundation.} to capture simultaneous spectra of Ca II and
Balmer lines in a sample of nearby M3 dwarfs. Our goal is to
investigate the chromospheric temperature structure required to
produce these lines at the observed levels. We find a strong positive
correlation between instantaneous measurements of Ca II K and the
Balmer lines in active stars, although these lines may not be
positively correlated in time-resolved measurements.  The relationship
between H$\alpha$ and Ca II K remains ambiguous for weak and
intermediate activity stars, with H$\alpha$ absorption corresponding
to a range of Ca II K emission. A similar relationship is also
observed between Ca II K and the higher order Balmer lines. As our
sample consists of a single spectral type, correlations between these
important chromospheric tracers cannot be ascribed to continuum effects, as suggested by other authors. 
These data confirm prior non-simultaneous observations of 
the H$\alpha$ line behavior with increasing activity, showing an
initial increase in the H$\alpha$ absorption with increasing Ca II K
emission, prior to H$\alpha$ filling in and eventually becoming a pure emission
line in the most active stars. We also compare our optical
measurements with archival UV and X-ray measurements, finding a
positive correlation between the chromospheric and coronal emission
for both high and intermediate activity stars. We compare our results
with previous determinations of the active fraction of low mass stars, and
discuss them in the context of surface inhomogeneity. Lastly, we
discuss the application of these data as empirical constraints on new
static models of quiescent M dwarf atmospheres.
\end{abstract}

\keywords{stars: late-type --- stars: low-mass, brown dwarfs ---
methods: data analysis}

\section{Introduction}

Radiation from the chromospheres of Sun-like stars comes primarily
from the Ca II resonance lines in the blue and Ca II infrared triplet, the Fe
II and Mg II resonance lines in the near UV, and the Balmer series of
Hydrogen in the optical \citep{linsky1982,foukal1990,nlds}. In M
dwarfs, high surface gravity compresses the stellar atmosphere,
leading to higher density chromospheres and particularly strong Balmer
emission \citep{linsky1982}. While the Ca II lines are the brightest
emission lines formed in the of the Solar chromosphere, contributing a radiative loss of
$\sim$ 5$\times$10$^6$ ergs cm$^{-2}$ s$^{-1}$ \citep{foukal1990}, the
Balmer emission in M dwarfs is much larger than all other
chromospheric lines, comprising $\sim$10$^{-4}$ of the bolometric
luminosity at early types \citep{linsky1982,we04}. This fact, combined
with the intrinsic faintness of M dwarfs in the blue, has made the red
H$\alpha$ line the primary diagnostic spectral feature of
chromospheric activity in these stars. Many M dwarfs also have
prominent Ca II resonance line emission-- particularly Ca II K, which
is unblended even at low resolution (as opposed to Ca II H, which is
blended with H$\epsilon$). The subset of these stars which also show
H$\alpha$ emission are designated ``active'' (dMe).

H$\alpha$ and Ca II K are two of the strongest optical emission lines
in active M dwarf chromospheres, responsible for cooling the
atmosphere, balancing the magnetic heating, and determining the
resulting equilibrium structure \citep{sund88,mf94}. Across the M
spectral class, there is a range of emission strength in Ca II K, and
a wide variety of both absorption and emission in
H$\alpha$. H$\alpha$ appears to trace hotter regions of the
chromosphere ($\ge$ 7000 K), while Ca II K traces the cooler regions
between the temperature minimum and $\sim$ 6000 K
\citep{gia82,cm85,walk09}. Thus together, H$\alpha$ and Ca II K offer
complementary information on chromospheric structure.

The nature and relationship of the Ca II K and H$\alpha$ lines in both
active and inactive M dwarfs has been controversial. It may be that Ca
II K and H$\alpha$ are not only complementary but integral pieces of
information, which separately cannot offer a true understanding of the
processes they trace. One point of contention has been the ``zero
point'' of chromospheric activity, or what defines the weakest
existing chromospheres. \citet{1984ApJ...282..683Y} have argued that
the zero point  for chromospheric activity is defined by strong
H$\alpha$ absorption. In contrast, \citet{sh86} claim that those stars
with the weakest chromospheres also have the weakest H$\alpha$
absorption. In their scenario, H$\alpha$ begins as a weak, purely
photospheric absorption line, which (when the atmosphere is subject to
magnetic heating) first increases in absorption equivalent width, then
decreases as it fills in with emission, then becomes a pure emission
line. The implication of this scenario is that both stars with
intrinsically weak chromospheres, and stars with moderate amounts of
magnetic activity, may exhibit weak H$\alpha$ absorption. Without
additional chromospheric indicators, such as the Ca II K line, it
is impossible to distinguish intermediate activity stars from inactive
stars. In M dwarfs, the cold temperature minimum is largely
transparent to H$\alpha$ photons \citep{pet81}, so only a very weak
H$\alpha$ absorption line is produced in the photosphere
\citep{cm85}-- thus we adopt those stars having weak H$\alpha$
absorption ($\le$ 0.1\AA) and weak Ca II K emission ($\le$ 0.1\AA) as
being representative of the ``zero point'' of activity in our sample.

Prior studies of the relationship between H$\alpha$ and Ca II K
\citep{rcg90,2006PASP..118..617R,cdm07} have
routinely observed a positive correlation between these lines in
active M dwarfs. However, both \citet{rcg90} and \citet{cdm07} have
suggested that this observed correlation is due to a color effect-- in
other words, that the relationship between these lines observed in a
sample of mixed spectral types is due to enhanced contribution from
high continuum fluxes in earlier type stars. These authors argue that
because time-resolved observations of Ca II K and H$\alpha$ emission
are not always positively correlated in individual stars, the positive
correlation between instantaneous measurements of these lines is the
result of their spectral type, rather than the properties of their
chromospheres.

Observations of Ca II K and H$\alpha$ provide an important ``point of
contact'' in linking theory to actual M dwarfs. Previous efforts to
study this relationship have often relied on observations from
different epochs to provide measurements of both lines for a given
star \citep{cram87,hs97}. In active M dwarfs, however, both Ca II K
and H$\alpha$ vary on timescales from minutes to decades. Therefore,
measurements of lines obtained from non$-$simultaneous observations
are not truly comparable, and any attempt to  model them may shed no
greater light than comparing lines in two totally different
stars. Indeed  nearly every previous study, observational and
theoretical, cites  non$-$simultaneous data as the chief source of
uncertainty in observed  relations, and recommends synoptic
observations of a sample of M dwarfs \citep{cram87,rcg90}.

Although many surveys have measured the Balmer lines of low mass
stars, particularly H$\alpha$ \citep[e.g. the Palomar-Michigan State
  University Survey, or PMSU;][]{gi02}, the desired
dataset of synoptic Balmer and Ca II K observations is sparse. The
non-simultaneous observations taken in previous surveys, along with
samples that span a wide range of M spectral types (and therefore a
wide range in color and absolute magnitude), have left the
relationship between these important chromospheric tracers ambiguous.
To remedy this uncertainty, we have carried out an extensive observing
program of H$\alpha$ and Ca II K in nearby M3 dwarfs. By taking
simultaneous data for a single spectral class, we are able to
determine if the observed correlation between these chromospheric
tracers is due to a continuum effect, or to an actual relationship
between these lines in active chromospheres. We also examine whether
multiple observations of the Balmer and Ca II lines in an individual
star are correlated over time in the same sense as single
observations, and if not, how this behavior can be reconciled with
instantaneous measurements of these lines in single observations of
that same star.

We also compare these observations with previous determinations
of the active fraction of low mass stars \citep{we04,we08}, which
have relied solely on the H$\alpha$ line as an activity proxy,
although it has been widely assumed that most M dwarfs possess some
level of Ca II K emission. Using the Ca II K line as an additional
activity measure, this survey will allow us to determine what fraction
of stars deemed ``inactive'' on the basis of H$\alpha$ absorption
actually do have Ca II K emission and therefore belong to an
intermediate activity population.

Additionally, we examine the proposed H$\alpha$ behavior
found by \citet{sh86}, whereby H$\alpha$ first increases in absorption
depth prior to filling in and going into emission. The \citet{sh86}
scenario was derived from non-simultaneous observations of a sample
composed of cool stars of a relatively wide range of
mass. \citet{cm85} reported some success reproducing these
observations using very simple model chromospheres, consisting of a
slab with a constant temperature and density. Using this simple model,
\citet{cm85} predict a different slope for the filling in of the
H$\alpha$ line depending on whether the increase in equivalent width
is due to an increased heating rate in a homogeneous chromosphere, or
an increased filling factor of active regions in a laterally
inhomogeneous chromosphere. If the filling in of the H$\alpha$ line is
evident in our observations, it will provide further
constraints on how chromospheric structure changes with increased
activity, as well as the role played by lateral inhomogeneity.

We are motivated to investigate surface inhomogeneity by the results
of \citet{we08}, who used the observed fraction of active stars per
spectral type, along with a simple dynamical model of the Galaxy, to
infer activity lifetimes for M dwarfs across the spectral class. A
large increase in the activity lifetime occurs over the range of
M3-M5, corresponding to the mass boundary where stars are believed to
become fully convective \citep[$\sim$0.35M$_{\odot}$;][]{nlds,cb97}. The transition to full convection may mark a
change from a Solar-type, rotationally-dependent magnetic dynamo, to
a turbulent dynamo that may or may not depend strongly on rotation. The decline
in activity in a given star over its lifetime may also indicate a
change in how that star's activity is generated: objects of
transitional mass (such as the M3 stars we examine here) may still
possess some marginal radiative zone, though the majority of the star
is characterized by deep convection. In this case, it is possible that
during the star's youth the activity is dominated by a Solar-type
dynamo, where the boundary of the convective and radiative zones is
responsible for most of the field generation, though a turbulent
dynamo is also at work. As the star spins down with age, the
rotation-dependent $\alpha\Omega$ dynamo declines, and the turbulent
dynamo becomes the predominant means of field generation. Indeed,
recent simulations of the magnetic dynamo in fully convective stars by
\citet{bro07} show that the turbulence-generated field has both large-
and small-scale features. The typical size of these field structures
are related to the radius, with those generated deeper in the interior
being larger than those near the surface. In addition, the large scale
field in these fully convective models can be generated and sustained
even though differential rotation (which plays such a key role in
Solar-type dynamos) is eventually quenched by the field
itself. Interestingly, although the \citet{bro07} models are
convective throughout, convection in the core is relatively weak, and
most of the energy is transported via the radiation.

There is some precedence for this ``dual-dynamo'' scenario on the Sun,
which possesses both a large-scale global field and a small scale
magnetic component that seems to be independent of the global
field. The gross magnetic features of the Sun are thought to be due to
the $\alpha\Omega$ dynamo that originates in the tachocline, but the
magnetic field of the quiet Sun is likely generated by turbulent
convective motions close to the stellar surface
\citep{men86,dur93,vo07}. The chromosphere is a thin layer, dynamic
but not convective, threaded by convection-twisted magnetic fields
that form the quiet Sun magnetic network \citep{ab08}. Field
generation by turbulence near the surface is not to be confused with
theories of $\alpha^2$ dynamos, which generate global scale turbulent
fields, or with acoustic heating of the chromosphere. Although
acoustic heating is bound to be at work wherever convective motion is
involved, purely mechanical means apparently cannot fully account for
the chromospheric and coronal emission in the Sun, and the heating due
to these surface turbulent fields is most likely magnetic in nature
\citep{ber05,ber06}.

In the stellar atmosphere, the manifestation of this behavior may be
surface inhomogeneity, where the chromosphere can be represented by
two components: one corresponding to magnetically active plage regions related to the
large-scale field, analogous to active regions on the Sun, the other
component representing a basal, turbulence-driven
chromosphere \citep{schrijver1989}. Observations of the solar surface
indicate that these
inhomogeneities are complex indeed-- Ca II K core emission corresponds
spatially to regions of concentrated magnetic field, such as in active
plage regions and bright network grains, while H$\alpha$ absorption is produced
in the both the mass of fibrils protruding from active regions, in
mottles across the network of the quiet Sun and in enhanced emission
from bright points during flares \citep{hasan,rutten06,rutten07}. The
diversity of structures in the solar chromosphere seems to indicate
that even a dual component description is somewhat simplistic in
nature-- the magnetic features of a given star likely form a more
continuous distribution, where the distinction between active and
quiet areas is unclear \citep{schrijver1988}. In the stellar case,
however, we have no such detailed observations of surface
inhomogeneities and must proceed cautiously with the number of free
parameters we allow in our interpretation of spatially averaged
spectra. However, if the Ca II or Balmer lines preferentially trace
particular chromospheric components, examining their relationship may
shed light on the nature of magnetic structures and the stellar dynamo
in convective-boundary mass stars.

In the next section, we describe the observed sample.In Section 3, we
discuss the observations and analysis, with particular attention to
the effect of spectral resolution on interpreting relationships
between chromospheric lines. In Section 4, we present the observed
relationships between Ca II and the Balmer lines. We also place these
observations in the context of previous surveys, addressing their
implications for chromospheric structure. Lastly, we comment on our
proposed theory of atmospheric inhomogeneity in light of the
observations, and discuss future plans to test this proposition with a
new suite of non-LTE model atmospheres.

\section{The Sample}

Our sample consists of 81 stars in the Palomar/MSU Nearby Star
Spectroscopic Survey \citep[PMSU;][]{re95a,ha96,gi02,re02} of the same
approximate spectral type, M3V (see Table 1 for a summary of our
sample stars). The PMSU Survey determined spectral types based on a
set of spectroscopic indices for TiO and CaH \citep[see][for details]{re95a}. Our sample is also quite homogeneous in color, having a
median Cousins V-I = 2.55 $\pm$ 0.12 (corresponding to a
difference in T$_{eff}$ of roughly $\pm$ 50K; \citet{nlds}). The
PMSU Survey H$\alpha$ measurements enabled us to select a sample that
is representative of varying levels of chromospheric activity, from
very weak H$\alpha$ absorption (indicating either a weak or
intermediate strength chromosphere), to powerful emission (indicating
strong chromospheric activity). Limiting the spectral range of our
objects is important, as it has been been noted \citep{sh86,ru87} that
the relation  between the chromospheric flux in the Ca II K line and
the equivalent width in H$\alpha$ depends on the spectral type of the
star (i.e. the stellar continuum flux). Previous studies have
attempted to correct for this effect by binning their measurements by
stellar color \citep{rcg90}.  By limiting our sample to a single
spectral class, we are able to examine the relation between Ca II K
and H$\alpha$ without having to disentangle systematic trends with
color and spectral type. Although many of the stars we observe have
existing optical data, our new data comprise simultaneous observations
of both lines at comparable resolution.

\section{Observations and Analysis}

\begin{figure}
  \begin{center}
\subfigure[Echelle spectra]{\label{fig:edge-a}\includegraphics[width=0.45\textwidth]{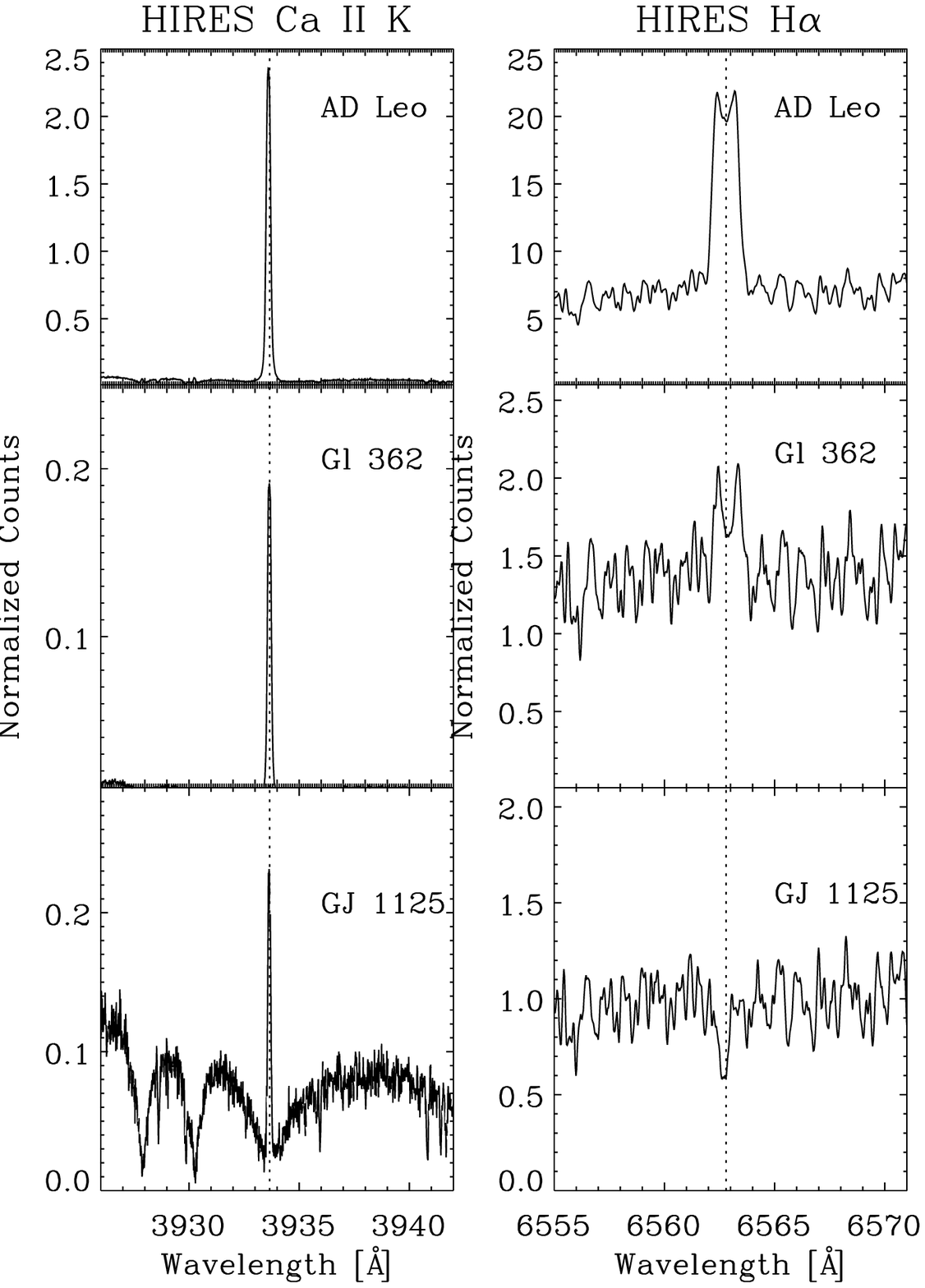}}
\subfigure[DIS Spectra]{\label{fig:edge-a}\includegraphics[width=0.45\textwidth]{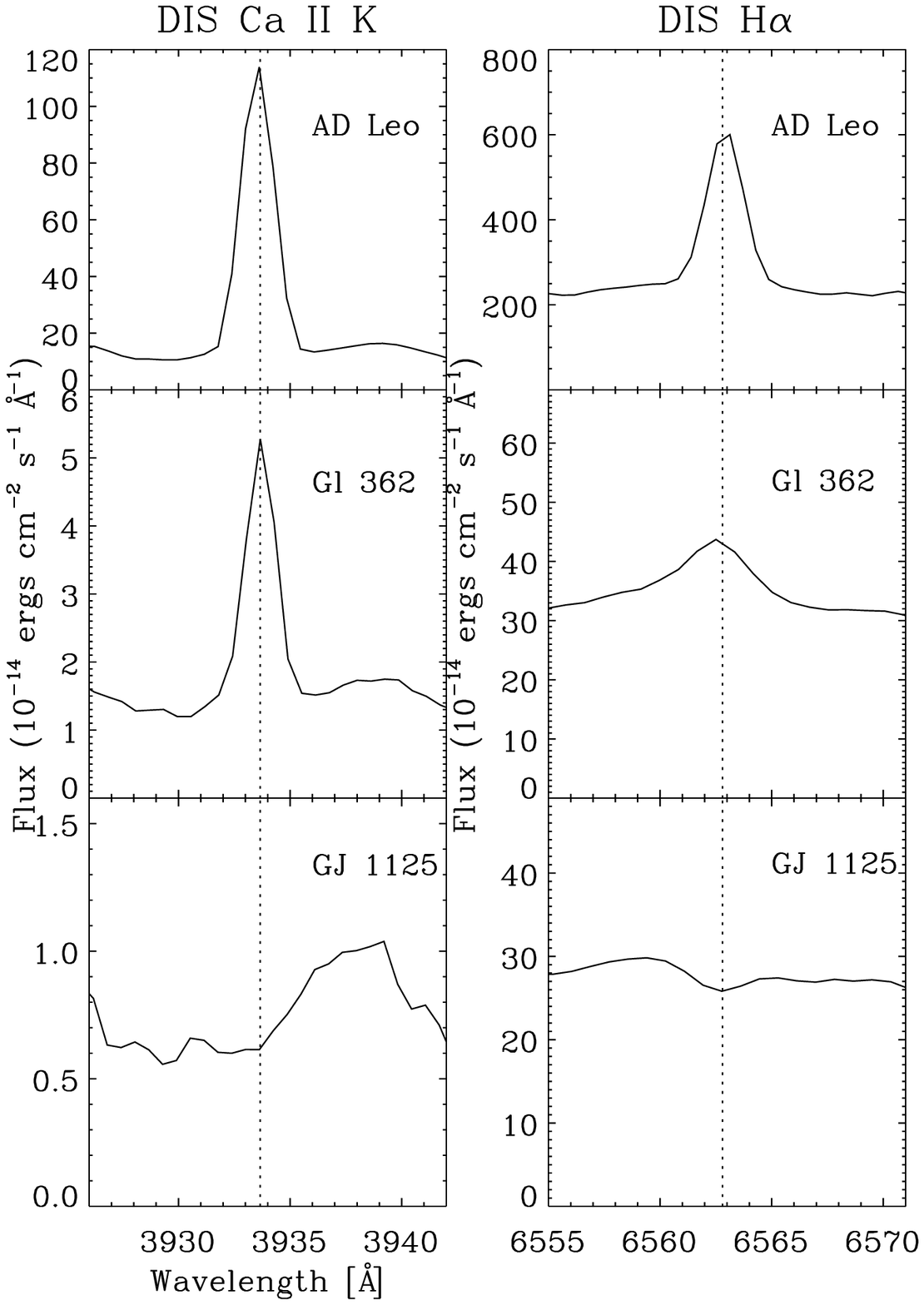}}
  \end{center}
\caption{Example HIRES (left) and DIS (right) observations of Ca II
  K and H$\alpha$ for three stars of varying activity
  level. Top: AD Leo, high activity; Middle: Gl 362, moderate
  activity; Bottom: GJ 1125, low activity. It is evident
  that even stars that appear to be inactive at low resolution
  (e.g. GJ 1125) may still show chromospheric Ca II K emission in high
  resolution spectra.}
\label{examplespec}
\end{figure}

Data were obtained with the ARC 3.5m Telescope at Apache Point
Observatory (APO), using the Dual Imaging
Spectrograph (DIS) and the ARC echelle spectrograph (ARCES). Additional high
resolution data (HIRES) were available for many of our stars from the
California Planet Search \citep[CPS;][]{bu96,wm04}. While the
DIS spectra provide reliable flux information for our lines of
interest, the echelle spectra enable an examination of the detailed
line profile. High resolution data are particularly important for
identifying intermediate activity objects, whose Ca II K chromospheric
emission cores may not be resolved in the lower resolution
spectra. Figure \ref{examplespec} shows example HIRES and DIS
spectra for three stars of different activity levels. In the highly
active star (top row: AD Leo), the H$\alpha$ and Ca II K lines are
quite clearly in emission in both the echelle and low resolution
spectra, indicating the presence of substantial chromospheric
heating. The moderate activity example (middle row: Gl 362) shows very
little H$\alpha$ emission, but still has a noticeable Ca II K emission
line. In the weakly active example (bottom row: GJ 1125), the
H$\alpha$ line is in absorption and Ca II K emission is not evident in
the low resolution spectrum. Based on its H$\alpha$ line, this star
would be classified as ``inactive''. However, the echelle spectrum of
this star illuminates the importance of high resolution observations--
even the ``inactive'' object possesses a noticeable Ca II K emission
core when examined in detail. In the following subsections, we
describe the observations, data reduction, and method of line
measurement for the DIS and echelle spectra. We also compare high and
low resolution measurements of the same stars, to investigate what
biases may be introduced by the choice of spectral resolution.

\subsection{DIS spectra}

DIS is a low
dispersion spectrograph with two cameras fed by a beam splitter,
allowing blue and red spectra to be obtained simultaneously. Our data
were taken using a 1\farcs{5} slit in combination with the higher
resolution gratings, which have a dispersion of 830.8 lines mm$^{-1}$
on the red side (giving a dispersion of 0.84\AA\ pixel$^{-1}$, R = $\lambda/\Delta\lambda$
$\sim$ 3200) and 1200 lines mm$^{-1}$ on the blue side (giving a
dispersion of 0.62\AA\ pixel$^{-1}$, R $\sim$ 2000). Spectra typically
covered the wavelength regions of 3700\AA\ to 4400\AA\ in the blue and
5600\AA\ to 7100\AA\ in the red.

The DIS data were reduced using standard IRAF reduction procedures \citep{iraf}. Data
were bias-subtracted and flat-fielded, and one-dimensional spectra
were extracted using the standard aperture extraction method. A
wavelength scale was determined for each target spectrum using HeNeAr
arc lamp calibrations. Flux standard stars from \citet{oke90} were
observed each night and used to place the DIS spectra on a calibrated
flux scale.

\subsection{Echelle Spectra}

The ARC echelle spectrograph (ARCES) captures the entire visible
spectrum (roughly 3800\AA\ to 9800\AA) with a resolution of
R$\sim$31,500. While this resolution is high enough that even small Ca
II K emission cores can be measured from our spectra, it does not
fully resolve the Ca II K line profile, and thus information regarding
the presence or lack of a central reversal in the detailed line
profile (an important tracer of NLTE effects in the upper
chromosphere) is lost in these spectra. Cosmic rays were removed from
the raw echelle images, which were then bias-subtracted and
flat-fielded. Scattered light in the ARCES spectra was removed to
the 6-8\% level using the apscatter task in the IRAF echelle
package, which fits a two-dimensional surface to the dispersed and
cross-dispersed interaperture light. One-dimensional spectra were extracted for data and ThAr arc
lamp images. The ThAr arc lamp spectra were then used to determine the
wavelength scale for each target spectrum using ThAr arc lamp spectra.

The CPS spectra were taken using the HIRES spectrograph on Keck I
\citep[R$\sim$57,000; ][]{vo94}, and cover the wavelength regime of
3645$-$7990\AA. The Keck HIRES data were reduced using the Planet
Search Reduction Pipeline, described in Wright et al. 2004 and
\citet{2006PASP..118..617R}. The pipeline removes scattered light from
the raw spectra by fitting a smooth function to counts lying between
the orders, and subtracting this level from the raw data. This
method removes the scattered light to 0 $\pm$ 1 photon (G. Marcy,
private communication). Equivalent widths for these objects have also
been reported in \citet{2006PASP..118..617R} and \citet{we06}, but we
have remeasured them as part of our analysis to ensure consistency.

\subsection{Measurement of Ca II K and Balmer Lines}

The observed line profiles are a superposition of the photospheric and
chromospheric contributions.  In the case of the Ca II K line, the
chromospheric emission core is easily identifiable in our echelle spectra even
for stars with low activity levels, and so the two components of the
line can be disentangled. As the distinction between
photospheric and chromospheric contribution to the Balmer lines is
considerably more ambiguous, we measure the entire
(photospheric$+$chromospheric) line. Measuring the entire H$\alpha$
line enables direct comparison between our work and previous
studies that use H$\alpha$ as an activity indicator, where
traditionally no distinction has been made between the chromospheric
and photospheric contributions to the line. 


The Ca II K line was measured in the same way for both the HIRES and
ARCES echelle data as well as the low resolution DIS data. As the Ca
II K line profile contains both contributions from the chromospheric
emission and the photospheric absorption, it is necessary to correct
for the photospheric contribution in order to study the chromospheric
emission itself. In order to remove the photospheric contribution to
each line, we use a radiative equilibrium NextGen model atmosphere
\citep[logg=5, T=3400K, solar metallicity;][]{phoenix} with the NLTE
radiative transfer code `RH' \citep{ui01} to calculate the
photospheric Ca II K line profile. The calculated line profile was
normalized and scaled to match the continuum near the Ca II K line for
each star. The model photospheric line profile was then overplotted
with the observed data, and a line integration region was
interactively chosen to include only the chromospheric emission; this
method is illustrated in Figure 4. As the stars in our sample are all
the same spectral type, the same model profile is used in each case
and the correction is applied in the same way to each spectrum.

The total area under the interactively chosen region is then
integrated both for the observed spectrum and the radiative
equilibrium model. The flux above the radiative equilibrium model is
taken as the chromospheric flux, and divided by the mean continuum level
measured from a relatively featureless portion of the nearby spectrum
to yield the equivalent width of the line (a summary of the wavelength
regions chosen for the mean continuum and line integration regions is
given in Table 2). We compare the
results obtained with photospheric correction for objects with both
echelle and DIS spectra in the next section. 

As the Balmer lines have relatively simple
profiles compared to the Ca II K line, such a detailed method of
measurement was not necessary. They were instead measured using an
automated method that sums the flux in rectangular passbands, using a
fixed integration region for each line. The Balmer
lines in the DIS spectra were measured using the same fixed
integration and continuum regions as for the high resolution
observations (specified in Table 2).

As the lower resolution DIS spectra are flux calibrated, fluxes for
the lines could be measured in addition to equivalent widths. Where
given, surface fluxes were calculated using distances from PMSU
\citep[derived from the M$_v$-TiO5 spectroscopic parallax relation, and
calibrated against those stars in the sample posessing trigonometric
parallaxes; see][]{gi02} and a typical radius for an M3V star
\citep[2.85$\times 10^{10}$ cm;][]{nlds}.

Table 3 provides the line fluxes and equivalent widths for
the entire sample as measured from the low resolution DIS spectra,
while Table 4 provides equivalent widths measured for the subset of
stars with echelle observations. 

\subsection{Comparison Between Low Resolution and Echelle Data}

The detailed profiles from our echelle spectra enable us to easily
disentangle the photospheric and chromospheric components of our
lines, which is particularly important for stars with low levels of
activity (and thus ambiguous low resolution spectra). We use the
comparison between the equivalents widths measured from the low and high
resolution spectra with photospheric correction to calibrate the
effect of spectral resolution on our results.

\begin{figure}
  \begin{center}
    \subfigure[All
    stars]{\label{}\includegraphics[width=0.45\textwidth]{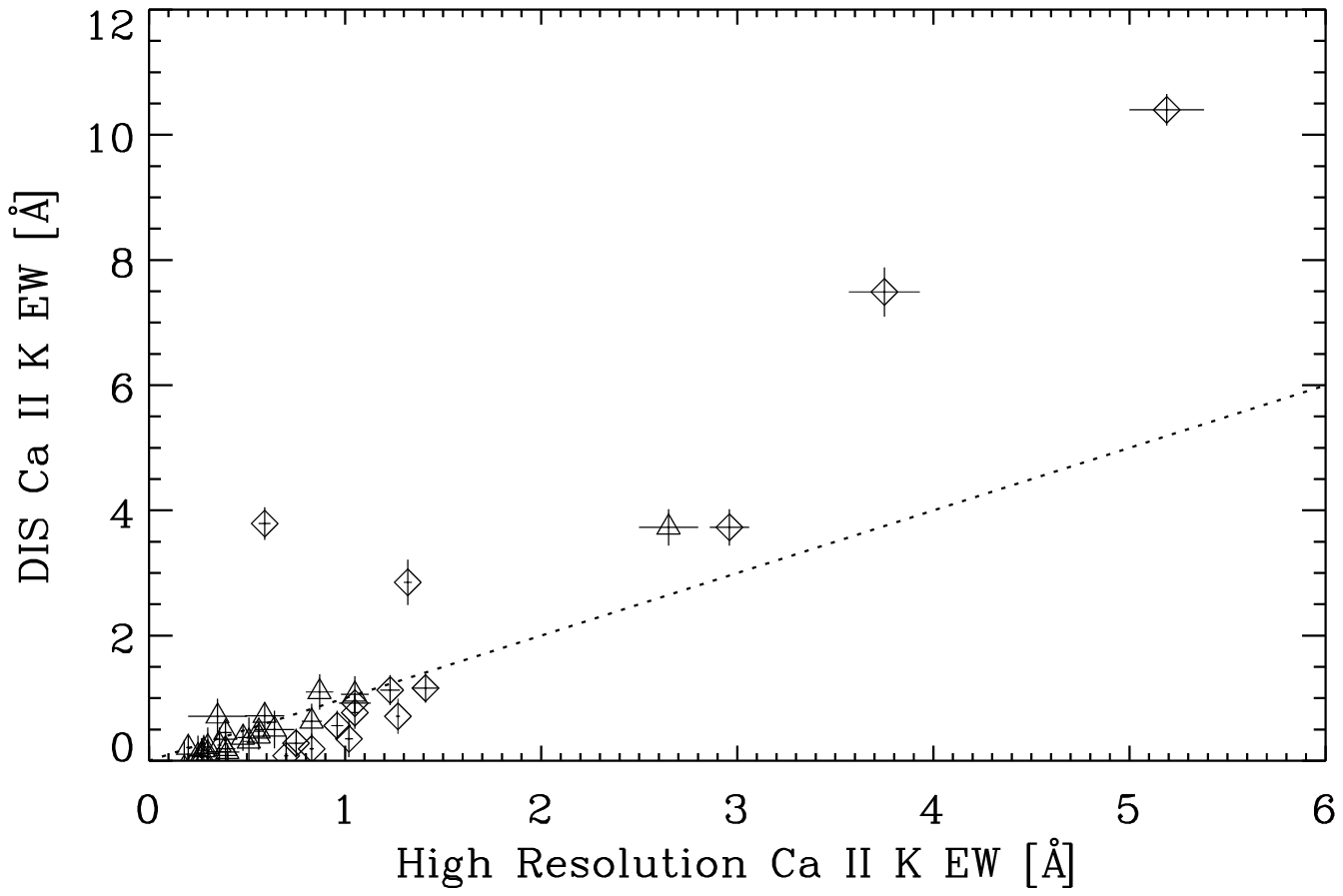}}
    \subfigure[Expanded scale of low activity
    objects]{\label{}\includegraphics[width=0.45\textwidth]{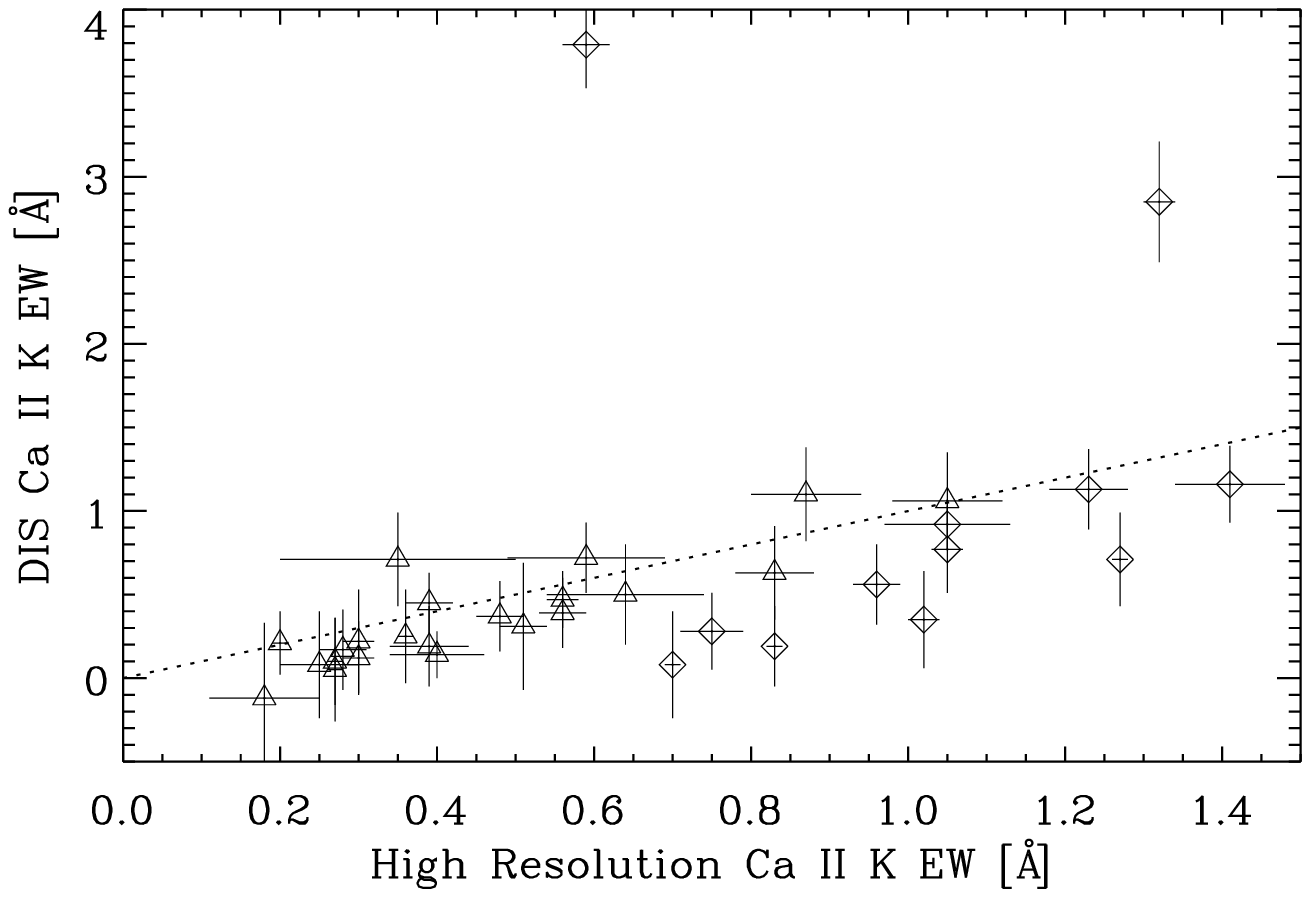}}
  \end{center}
\caption{Comparison between Ca II K equivalent widths measured from
low resolution data corrected for the photospheric contribution, and
equivalent widths measured by fitting the Ca II K emission core in
high resolution data. Diamonds denote measurements of our ARCES echelle data, while triangles indicate measurements from HIRES echelle data. The dotted line represents the locus of
agreement for the two equivalent width measurements. For small
equivalent width, the two methods of measurement produce comparable
results, and the points lie along the line of agreement. However, for
the most active the equivalent widths measured from the low resolution
data are larger than those measured from the high resolution
data. This difference is due to how the photospheric absorption line
is accounted for in our two measurement methods: the method used for
measuring the high resolution data underestimates the chromospheric
contribution to the line. The two points in (b) which lie
significantly above the locus are two stars that were observed with
DIS during higher activity states than their echelle observations.}
\label{disvechelleew_cak}
\end{figure}

\begin{figure}
  \begin{center}
    \subfigure[All
    stars]{\label{}\includegraphics[width=0.45\textwidth]{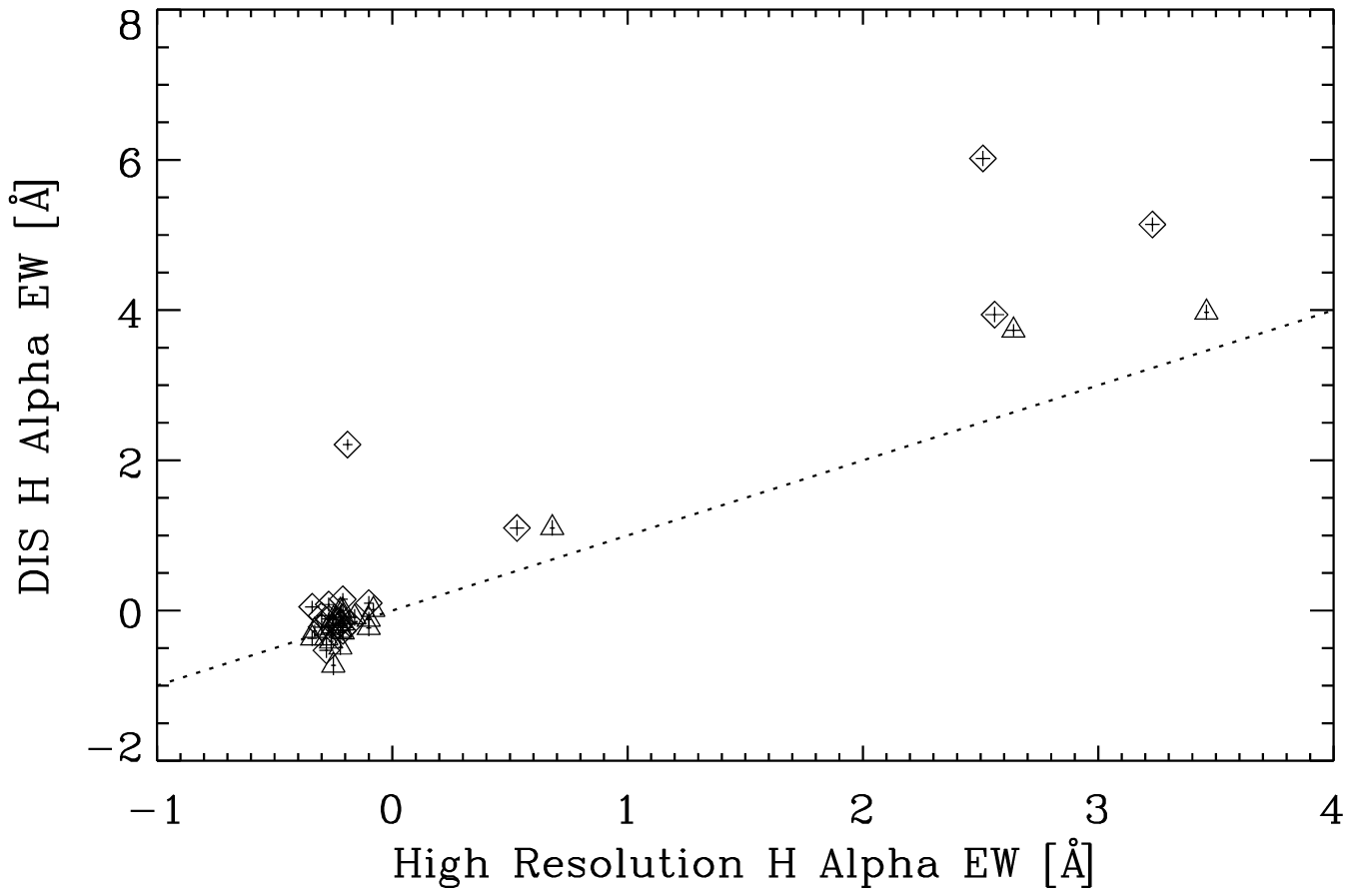}}
    \subfigure[Expanded scale of low activity
    objects]{\label{}\includegraphics[width=0.45\textwidth]{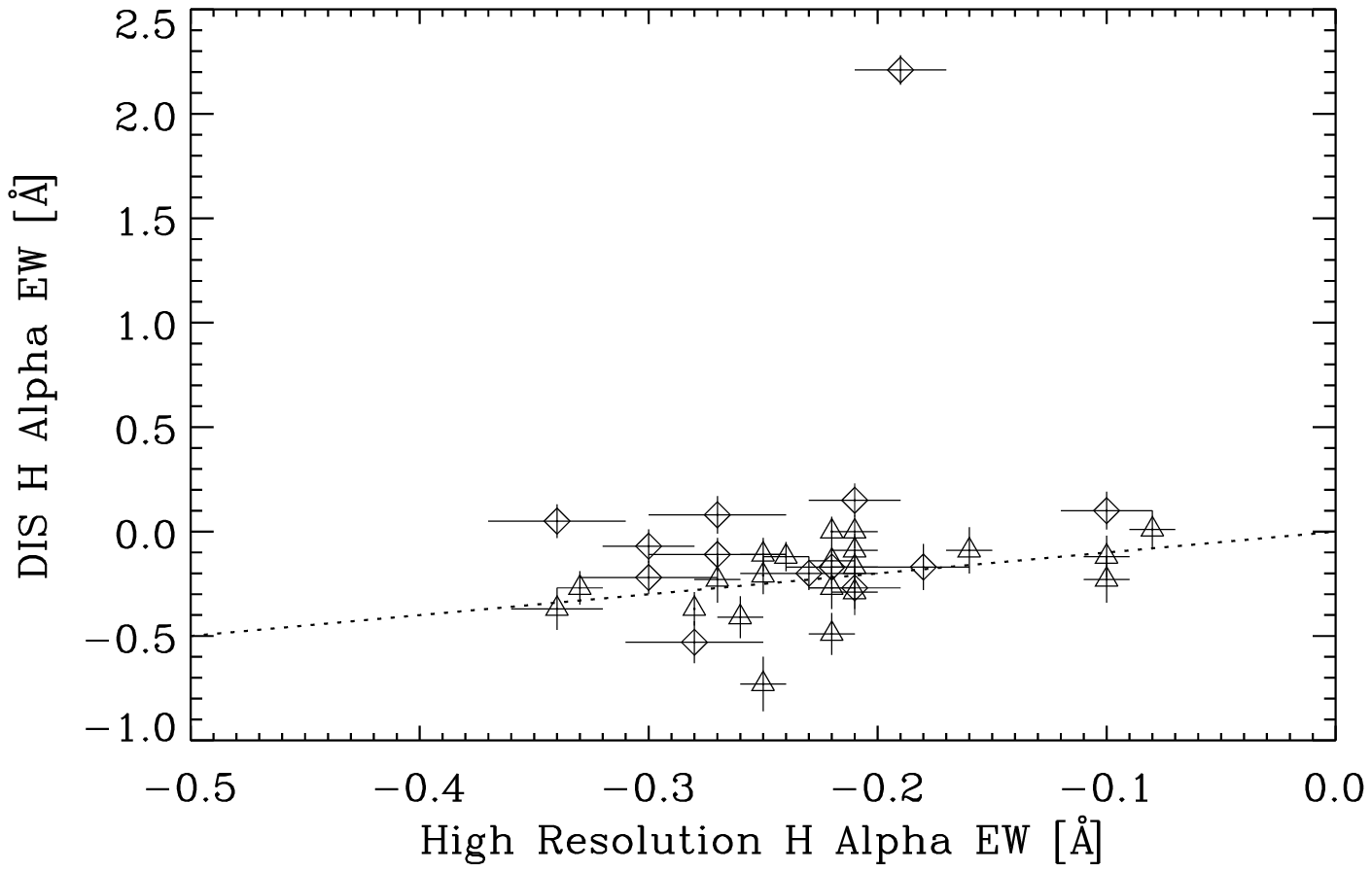}}
  \end{center}
\caption{Comparison between H$\alpha$ equivalent widths calculated
  from DIS and echelle spectra. The dotted line represents the locus
  of agreement for the two equivalent width measurements. Diamonds denote measurements of our ARCES echelle data, while triangles indicate measurements of HIRES echelle data. As in the
  case of the Ca II K line, the two methods of measurement produce
  comparable results for small equivalent widths. However, for the
  most active the equivalent widths measured from the low resolution
  data are larger than those measured from the high resolution
  data. This discrepancy is caused by the H$\alpha$ line falling close
  to the edge of the order in our echelle spectra-- the continua in
  our echelle spectra have a different slope than the low resolution
  spectra, which causes the continuum value, and thus the equivalent
  width, to be underestimated in the echelle measurements.  The point
  in (b) which lies significantly above the locus is a star observed
  with DIS during a higher activity state than its echelle
  observation.}
\label{disvechelleew_ha}
\end{figure}

Figures \ref{disvechelleew_cak} and \ref{disvechelleew_ha} compare the
equivalent widths obtained from the DIS spectra with those from the
echelle spectra. For low levels of activity, the Ca II K equivalent
widths derived from the corrected low resolution data are a decent
match with those derived from the high resolution data. The difference
in equivalent widths calculated from the ARCES spectra versus the DIS
spectra are $\pm$0.70$\mbox{\AA}$ in H$\alpha$ and
$\pm$1.11$\mbox{\AA}$ in Ca II K, while equivalent widths measured
from the HIRES spectra are $\pm$0.17$\mbox{\AA}$ in H$\alpha$ and $\pm$0.18$\mbox{\AA}$ in
Ca II K. However, as the activity of the star increases,
equivalent widths calculated from the low resolution data are a good
deal higher in both Ca II K and H$\alpha$: in H$\alpha$, the
equivalent widths measured from ARCES spectra of active stars are
2.13$\mbox{\AA}$ lower than those measured from the low resolution
spectra, while HIRES measurements are 0.99$\mbox{\AA}$ lower. In Ca II
K the effect is more pronounced, with the equivalent widths measured
from ARCES and HIRES spectra being 1.63$\mbox{\AA}$ and
3.73$\mbox{\AA}$ lower, respectively.


As our echelle spectra are not flux calibrated, the continua have
subtle differences in slope compared to the continua of the DIS
spectra. While this effect is small for the majority of the sample,
which have small H$\alpha$ equivalent widths, it becomes much more
noticeable when the line is in emission and contains considerably more
flux. In the case of the HIRES spectra, H$\alpha$ falls very close to
the edge of the reddest order, and there is no adjacent order with
which to join it.  In both the HIRES and ARCES spectra, the equivalent
widths calculated from the echelle spectra are smaller than those
calculated from the DIS spectra.

The differences between the two datasets indicate that equivalent widths calculated from high resolution
data will be smaller for both the H$\alpha$ and Ca II K lines in
active stars. Therefore, measurements taken from the high resolution
spectra will be offset in their absolute values of the equivalent
widths for the same star. However, comparisons between the high
resolution measurements of H$\alpha$ and Ca II K will show the same
relationship as for the low resolution data due to the net effect
being in the same direction for both lines. If correction for the
photospheric component of the Ca II K line is not performed, however,
the activity level will always be underestimated, especially for low
activity objects whose emission cores are not resolved. In the
following section, we only compare data of comparable resolution.

\section{Results}

In this section we present our results and interpret them in response
to the questions raised in the Introduction.

\subsection{The Relationship Between H$\alpha$ and Ca II K with Increasing Activity}

Comparison of the Balmer and Ca II lines reveals a number of
interesting relationships. Figure \ref{havcak_corrected} plots the
equivalent width of H$\alpha$ versus the equivalent width of Ca II K
as measured from our DIS spectra. The Ca II K line has been corrected
for the photospheric component as described in Section 3, so the
equivalent widths cited here represent the purely chromospheric
contribution to the line. For the most active stars (those with
H$\alpha$ in emission, indicating strong chromospheric heating), the
H$\alpha$ and Ca II K equivalent widths are clearly correlated and
increase monotonically, with increased scatter for stronger line
strength. A linear fit to the H$\alpha$ and Ca II K equivalent widths
for the active stars yields an RMS deviation of $\pm$0.38$\mbox{\AA}$,
with stars of Ca II K equivalent width of less than
$\sim$5.35$\mbox{\AA}$\footnote{This value corresponds to the mean of
the Ca II K equivalent widths measured for active stars and is a
largely arbitrary choice for the division between ``active'' and
``very active''. Our result that the scatter increases with increasing
activity is fairly insensitive to this value, routinely yielding an
RMS deviation for the most active stars that is roughly twice that
for less active stars.}  having an RMS deviation of
$\pm$0.21$\mbox{\AA}$, while the most active stars (EW$_{CaII K}$
$\ge$ 5.35$\mbox{\AA}$) have an RMS deviation of
$\pm$0.56$\mbox{\AA}$. As these data are measured from a homogeneous
sample of stars of the same spectral type, and the observations of Ca
II K and H$\alpha$ are simultaneous, the increased scatter is likely
intrinsic and suggests a mild decoupling of the line behavior in the
strongest chromospheres. Stars possessing weak to intermediate
chromospheres cluster in the lower left corner of the plot, and are
spread over a wide range in Ca II K for a relatively small range in
H$\alpha$. These weakly active stars,  plotted at an expanded scale in
Figure \ref{havcak_corrected_zoom}, do not show any obvious
correlation between the H$\alpha$ and Ca II K equivalent widths. That
a range in Ca II K emission strength can be associated with either
deep or shallow H$\alpha$ absorption lines lends credence to the
\citet{sh86} scenario, where the photospheric H$\alpha$ absorption
line ``fills in'' with chromospheric emission as activity
increases. However, an initial deepening of H$\alpha$ absorption as Ca
II K increases, suggested by \citet{cram87} and \citet{sh86}, is not
obvious in these low resolution observations.

Figure \ref{havcak_corrected_zoom} illustrates the number of stars
that would traditionally be classified as ``inactive'' based on having
H$\alpha$ in absorption, which actually do possess some chromospheric
activity (indicated by their Ca II K emission). Of the 81 stars in our
sample, 46 stars were classified as inactive based on having H$\alpha$
in absorption; with the additional information provided by our Ca II K
measurements, we find that all but possibly 1 of these stars have at least a
low level of Ca II K emission. As discussed in the introduction, this
result implies that H$\alpha$-based measurements of the active
fraction in {\em early} M dwarfs (above the fully-convective mass
boundary) may trace what fraction of stars have a solar-type magnetic
component that relies on the presence of a radiative-convective zone boundary (the tachocline). However, as H$\alpha$ emission is not only present but
increasingly common in very low mass, fully convective M dwarfs which do not possess a tachocline,
H$\alpha$ in these stars may be associated with a global turbulent
dynamo and thus trace an altogether different source of magnetic
activity (as discussed in the Introduction).

Figure \ref{havcak_corrected_echelle} shows echelle equivalent widths
of the Ca II K chromospheric emission cores against the full
(photospheric+chromospheric) equivalent width of the H$\alpha$ line as
a comparison with Figure \ref{havcak_corrected}. Figure
\ref{havcak_corrected_echelle_ZOOM} provides a similar comparison to
the low activity stars as Figure \ref{havcak_corrected_zoom}. As in
Figure \ref{havcak_corrected}, Figure \ref{havcak_corrected_echelle}
shows a very small variation in H$\alpha$ absorption for a range of Ca
II K emission. Figure \ref{havcak_corrected_echelle_ZOOM} is
intriguing, however, as the U-shape locus of points does seem to
indicate an initial deepening of the H$\alpha$ absorption line prior
to its filling in and going into emission. This relationship was also
observed by \citet{2006PASP..118..617R} in their analysis of Ca II K
HIRES data and H$\alpha$ equivalent widths from \citet{gi02} for M
dwarfs of mixed spectral types, but the trend persists if we examine a
homogeneous sample. Arrows indicating the approximate placement of the
locus predicted by \citet{cram87} are shown with the data to guide the
eye. The maximum H$\alpha$ absorption equivalent width is
$\sim$-0.35\AA, which is consistent with the range observed by
\citet{sh86}. The maximum in H$\alpha$ absorption seems to occur at
roughly 0.4 - 0.5\AA\ Ca II K equivalent width, suggesting that stars
having Ca II K emission at this level and above have passed some
critical threshold for the onset of H$\alpha$ emission.

\subsection{Chromospheric Coupling: Surface Fluxes and Higher Order Balmer Lines}

In Figure \ref{haewvcaksurf}, the H$\alpha$ equivalent width is
plotted against the chromospheric Ca II K surface flux for our DIS
sample. For M stars with H$\alpha$ in absorption, a relatively small
variation in H$\alpha$ equivalent width corresponds to a wide range of
Ca II K emission, confirming the results of Figure
\ref{havcak_corrected} above. For the strongest chromospheres, the
decoupling of Ca II K and H$\alpha$ emission suggested by Figure
\ref{havcak_corrected} is much more evident, with strong Ca II K flux
being associated with a wide range of H$\alpha$ emission
strengths. These results confirm those of \citet{rcg90}, although
their analysis was complicated by the spread in spectral type of their
sample (early K to mid M). They similarly corrected the Ca II K
for the photospheric
contribution to the equivalent width, but used a solar model scaled to the various
temperatures of their sample stars.

Figure \ref{higherlines} shows the relationship between the equivalent
width of Ca II K and those of H$\delta$, H$\gamma$, H$\epsilon$ and
H8. As in the case of Figure \ref{havcak_corrected}, the equivalent
widths are calculated from single DIS observations of our sample
stars. Our results provide an interesting comparison to \citet{we06},
who reported that in multiple observations of the Balmer and Ca II
lines in a given star, the higher order Balmer lines and Ca II H and K
lines trace one another, while H$\alpha$ varies seemingly
randomly. From this plot, it is evident that the higher order Balmer
lines follow Ca II K in a similar way to H$\alpha$. In contrast,
Figure \ref{thesis_adleo_multiplot} \citep[reprinted here from][]{we08b} shows the Balmer and Ca II K line
equivalent widths measured from repeat echelle observations of AD
Leo. Confirming the observation of \citet{we06}, we see that while the
higher order Balmer lines vary little in relation to Ca II K
($\Delta$EW$_{H\delta}$=$\pm$0.16$\mbox{\AA}$; $\Delta$EW$_{H\gamma}$=$\pm$0.15$\mbox{\AA}$; $\Delta$EW$_{H\epsilon}$=$\pm$0.12$\mbox{\AA}$), the
H$\alpha$ equivalent width can vary dramatically over time ($\Delta$EW$_{H\alpha}$=$\pm$0.57$\mbox{\AA}$). Taken
together, Figures \ref{havcak_corrected}, \ref{higherlines}, and
\ref{thesis_adleo_multiplot} imply that while the emission in Ca II K
and the Balmer lines are positively correlated when measured from a
single observation of a given star, they are not necessarily
positively correlated over multiple observations of that star. In
other words, more active chromospheres produce greater emission in
both Ca II K and the Balmer lines overall, but H$\alpha$ and Ca II K do not
always trace one another directly in time-resolved measurements.

\subsection{Investigation of Continuum Effects}

Although our sample is comprised of stars of the same spectral type,
our stars cover a range of $\sim$2 magnitudes in M$_v$. As previous
studies by \citet{rcg90} and \citet{cdm07} have suggested that the
positive correlation between Ca II and the Balmer lines in active
stars is due to enhanced continua in early spectral types, we
investigated the possible relationship between absolute magnitude and
activity indicators for our sample. Figure \ref{mvcont} shows the
continuum fluxes near H$\alpha$ and Ca II K versus the absolute V
magnitude, while Figure \ref{mvlines} shows the equivalent widths of
H$\alpha$ and Ca II K versus absolute V magnitude. The absence of any
relationship between absolute magnitude and the level of either
activity or continuum flux shows that the positive correlation we
observe between H$\alpha$ and Ca II K in active stars cannot be
ascribed to a continuum effect.

\subsection{Interpretation}

A possible interpretation of these results is that the Ca II K and
H$\alpha$ fluxes observed are due to a laterally inhomogeneous
atmosphere. Stated simply, increased activity leads to stronger
emission in all lines, but does not imply that these lines trace the
same source of activity or the same regions on the star. In analogy to
the ``zebra effect'' suggested by \citet{zebra} to describe
photometric variation on BY Dra stars, one may picture the stellar
atmosphere as consisting mostly of a cooler `basal' chromosphere,
heated either acoustically, through a weak, pervasive magnetic field,
or some combination of the two. Stars with greater activity may have,
in addition to this basal chromosphere, hotter active regions where
magnetic flux tubes thread through the stellar atmosphere, analogous
to bright plage regions on the Sun. The stronger the stellar magnetic
field, the more area of the stellar surface is covered by these active
regions. The observed line profiles are a superposition of the line
profiles produced by these two atmospheric components.

The Ca II K emission core responds linearly with increased activity--
more active chromospheres produce stronger Ca II K emission. The
H$\alpha$ line, on the other hand, may be found in various degrees of
absorption for low to intermediate levels of activity, and in emission
for strong activity. Let us describe a few limiting cases: (1) the
majority of the stellar surface is covered with quiet chromosphere and
a small fraction of very active regions; (2) the majority of the
stellar atmosphere consists of the active component, with a small
fraction of the quiet chromosphere, and (3) the majority of the
surface is covered in quiet chromosphere and a small fraction of
moderately active regions. In the first case, as an active region on
the star rotates into view, the equivalent width of Ca II K emission
increases, while the H$\alpha$ absorption line fills in and also
increases in equivalent width (i.e. becomes a shallower and thus
``less negative'' absorption line). Once this isolated active region
rotates out of view, both lines decrease. Our first case would show a
positive correlation between Ca II K and H$\alpha$ equivalent widths
measured over time. In our second case, the majority of the star is
producing both Ca II K and H$\alpha$ in emission. As a region of quiet
chromosphere rotates into view, Ca II K decreases slightly as both the
basal and active Ca II K lines are in emission. The observed H$\alpha$
line is now a superposition of an absorption line with the
chromospheric emission line, causing the equivalent width to
decrease. Again, one would observe a positive correlation between Ca
II K and H$\alpha$. In our last scenario, the majority of the star is
covered by basal chromosphere producing a weak H$\alpha$ absorption
line and Ca II K emission, and a moderately active region comes into
view. If this moderately active region produces a deeper H$\alpha$
absorption line, the equivalent width of H$\alpha$ will decrease,
while the Ca II K emission will be enhanced.  In this case, one would
observe a negative correlation between Ca II K and H$\alpha$. In
between these extremes, variations in active region coverage and
distribution across the stellar surface may explain much of the
observed behavior between these two chromospheric tracers.
\section{Comparison to UV and X-Ray Data}

In order to investigate the relationship between the chromosphere and
the corona, we compare our optical measurements to archival near-UV and X-ray
data. Of the 81 stars observed, 30 have archival X-ray measurements,
and a subset of 7 of these stars have near UV (Mg II)
measurements. All UV measurements come from active stars (those with
H$\alpha$ in emission), while the X-ray subsample also includes some
moderate and weak activity stars (which have H$\alpha$ in
absorption). In Table 5, we provide previously published Mg II and
X-ray measurements \citep{wa08,bo85,jo86,md89,md92} where available.

Figure \ref{xrayvmg} plots the X-ray flux versus the Mg II flux for
this subsample. Both the X-ray and UV observations were drawn from
various sources: X-ray values were taken either from ROSAT
observations of nearby stars (Huensch 1999) or EINSTEIN/EXOSAT values
originally reported in \citet{md89}, noted in the plot legends as
``MD89''. Measurements of the Mg II flux were taken from either IUE
observations \citep{md89} or ACS observations \citep{wa08}. As some of
the stars in this subsample have been observed by multiple
instruments, the points in each plot are color coded by where the
observations originate, and multiple observations of a single object
are connected by solid lines between points. Although a number of M
dwarfs have been observed with newer X-ray instruments (such as XMM
and Chandra) and with higher resolution in the UV (as with STIS or
GHRS), our sample is comprised only of M3 dwarfs. We include the ROSAT
and \citet{md89} data sets because they are available for a number of
stars in our sample, not just a select few of the most active ones
(e.g., AD Leo and EV Lac). The solid line located to the right is the
approximate power law fit to the Mg II and soft X-ray emission in G
and K dwarfs reported in \citet{ayres81}, while the dashed line
represents equality between the Mg II and X-ray emission. Our sample
seems to obey a similar power law relation as that found by
\citet{ayres81} for earlier type dwarfs.

Our X-ray and Mg II detected subsample can also be compared to the
H$\alpha$ and Ca II K measurements reported here, as shown in Figures
\ref{xrayvopt} and \ref{mgvopt}, respectively. In the plots comparing
the flux in the H$\alpha$ line, filled points indicate H$\alpha$ in
emission and unfilled points indicate H$\alpha$ in absorption. The
Ca II K line always appears in emission.

Figure \ref{mgvopt} shows archival UV measurements of Mg II versus our
measurements of H$\alpha$ and Ca II K. As the Mg II observations are
all of relatively active M dwarfs, this figure shows a positive
correlation between Mg II and both H$\alpha$ and Ca II K, reminiscent
of the relationship between H$\alpha$ and Ca II K in active stars shown
in Figures 4a and 5a. 

Figure \ref{xrayvopt} shows a clear correlation between the optical
chromospheric emission and the coronal X-ray emission. The
relationship between the coronal emission and the H$\alpha$ emission
has somewhat less scatter than that between the X-ray and Ca II K flux
(the three points that sit to the left of the main locus in Figure
\ref{xrayvopt}a, which appear to have anomalously low H$\alpha$ flux,
are probably intermediate activity stars whose H$\alpha$ lines are
almost completely filled in with chromospheric emission). The
relationship between H$\alpha$ and X-rays may be indicative of the
link between emission from the upper chromosphere (where H$\alpha$ is
formed) and the coronal emission. Figures \ref{xrayvmg} and
\ref{xrayvopt} show that the Ca II K, H$\alpha$ and Mg II chromospheric lines trace the
coronal X-ray emission in the same way as one another, likely due to this
subsample being comprised of active stars, whose greater activity
leads to more emission in all lines.

Plasma temperature distributions have been calculated using Chandra
data for only two of the
most active stars in our sample, AD Leo and EV Lac; the bulk of
the plasma lies between 2$-$20 $\times$ 10$^6$ K, peaking around 7$-$8
$\times$ 10$^6$ K for both stars \citep{rs05}. However, ROSAT hardness
ratios, which compare the flux in the `soft' (0.14 $-$ 0.42 keV; `S') and
`hard' (0.52 $-$ 2.01 keV; `H') bands, are available for a larger number of
stars. The hardness ratio, (H$-$S)/(H$+$S), can be used as a rough
proxy for coronal temperature, where harder emission (larger hardness
ratio) corresponds to hotter coronal plasma.  Figure
\ref{hardnessvopt} shows the hardness ratio for ROSAT detected stars
versus the flux in Ca II K and H$\alpha$. It is evident from both of
these plots that the chromospheric emission, as traced by H$\alpha$ is
positively correlated with the hardness of the coronal emission,
albeit with large scatter. In the case of Figure \ref{hardnessvopt}b,
the two points that appear to have low values of H$\alpha$ flux for
their coronal hardness are again the intermediate activity stars,
where the equivalent width of the line is very small due to its being
mostly filled in with emission. The Ca II K chromospheric emission
also appears to be loosely correlated with the hardness of the coronal
emission, but with far greater ambiguity than seen in the case of
H$\alpha$.

We can further examine the relationship between the coronal and
chromospheric emission by looking at the ratio of the luminosity in
H$\alpha$ to the X-ray luminosity versus the hardness ratio. In Figure
\ref{hardness}, unfilled symbols indicate H$\alpha$ in absorption,
while filled symbols indicate H$\alpha$ in emission. The two points
that lie below this locus are the same two stars with very small
H$\alpha$ fluxes due to the line being completely filled by
chromospheric emission. The median value for the sample as a whole is
log(L$_{H\alpha}$/L$_x$) $\sim$ $-$0.6, which is consist with the
values reported by \citet{re95b} for low mass stars in the Hyades
(log(L$_{H\alpha}$/L$_x$) $\sim$ $-$0.6) and Pleides
(log(L$_{H\alpha}$/L$_x$) $\sim$ $-$0.8). For the most part, there is
no obvious trend in the data-- however for the most active stars,
there appears to be a slight indication that the X-ray luminosity
increases proportionately with the coronal temperature, shown here as
a slight negative correlation in the filled data points. Considering
the dM and dMe stars separately, the median value of
log(L$_{H\alpha}$/L$_x$) for the dM stars is log(L$_{H\alpha}$/L$_x$)
$\sim$ $-$0.45, somewhat higher than  the dMe sample, where
log(L$_{H\alpha}$/L$_x$) $\sim$ $-$0.84. These
results are also consistent with those reported in \citet{sku84} and
\citet{flem88} for field M dwarfs.

Figure \ref{hardness} also compares the ratio of the chromospheric to
coronal emission, respectively measured by H$\alpha$ and X-rays, to
model predictions of the minimum X-ray flux in late type stars. The
colored bars in this plot correspond to our measured H$\alpha$ flux
divided by the range of X-ray flux generated by various models of
coronal heating: model predictions of acoustically heated coronae
\citep[log F$_x$ = 5.1 ergs s$^{-1}$ cm$^{-2}$;][]{mc94}, and the
X-ray flux predicted by models of magnetic field generation in the
convective envelopes of non-rotating stars \citep[log F$_x$ = 3.8 ergs
s$^{-1}$ cm$^{-2}$;][]{ber05}. Red bars indicate stars with H$\alpha$
in emission, while blue bars indicate stars with H$\alpha$ in
absorption. Points lying below their accompanying bars {\em exceed}
the amount of coronal emission that can be explained by either of
these models, while the X-ray flux of points lying within their
colored bars could potentially be explained by either prediction. All of the active stars (dMe; filled diamonds)
exceed the level of coronal emission produced in purely acoustic
models, in accordance with the role global, rotationally dependent
magnetic activity is believed to play in the heating of the outer
atmospheres of these stars. In the case of the dM stars (open
diamonds), the behavior is more complicated-- in most cases the
coronal flux exceeds both predictions, but in some cases it does
not. This observation may provide another indication that two
components are at work in the outer atmospheres of M dwarfs-- hot,
active regions associated with global magnetic activity, coexisting
with cooler basal regions that may be associated with small
scale convectively-generated magnetic fields, analogous to quiet regions on the Sun. The
relative filling factors of these two components may explain much of
the variation we see in M dwarf chromospheric and coronal emission.

\section{Conclusions and Future Work}

The observations presented in this paper provide a rich set of
empirical constraints for the construction of new model
chromospheres. We find that for weak to intermediate activity stars,
there is a range in the equivalent width of Ca II K emission for a
small range of H$\alpha$ absorption. Our high resolution data indicate
that the H$\alpha$ line may undergo an initial increase in absorption,
as Ca II K increases, with the line filling in and eventually going
into emission for the most active stars as observed by \citet{sh86}
and predicted by \citet{cram87}. From H$\alpha$ observations alone, it
is impossible to distinguish between a moderately active and an
inactive star. Therefore, studies of activity in low mass stars that
focus solely on stars with H$\alpha$ in emission are biased towards
the most active chromospheres. These active stars have both Ca II
and Balmer lines in emission, and more active stars have stronger
emission in both lines. This positive correlation between Ca II K and
Balmer emission in single observations of active stars does not imply,
however, that the chromospheric tracers are positively correlated
over time for every star. The variation in chromospheric emission of
individual stars is most likely a function of the quiet and active
region coverage unique to their particular stellar atmosphere.

That lateral inhomogeneity may play a major role in the observed
spectra of our M3 sample is not surprising, given observations of active
regions on the Sun, and the inferred large filling factors of active
regions deduced from M dwarf magnetic field measurements and
observations of flares \citep[e.g. ][]{basri90,cjk96,oneal04}. In the
scenario we propose, stronger chromospheric emission would correspond
to a greater presence of active regions on the stellar surface due to
the increased influence of the magnetic field. If Ca II K and
H$\alpha$ are produced preferentially by cool and hot components,
respectively, one would expect that while greater chromospheric
heating would cause more emission in both lines, the relation between
the two lines would be increasingly decoupled as the hotter active
component becomes more prominent.

We are in the process of further investigating whether two-component
model atmospheres can account for the relationships between the Ca II
and the Balmer lines. Using the data described here as empirical
constraints, we are developing a new generation of quiescent non-LTE
model atmospheres for M dwarfs. These two-component static models will
be calculated using the non-LTE radiative transfer program `RH'
described in \citet{ui01}. RH allows for bound-bound and bound-free
radiative transitions to overlap in wavelength and includes the
effects of partial redistribution for strong bound-bound transitions,
essential for proper treatment of the Ca II and Balmer lines. This
effort complements recent work on modeling the outer atmospheres of M
dwarfs during flaring states with the radiative-hydrodynamic non-LTE
transfer code RADYN (Carlsson \& Stein 1997; Abbett \& Hawley 1999;
Allred et al. 2005). These new models will address both the observed
line fluxes and correlations between Ca II K and the Balmer lines
obtained from our lower resolution data, and in addition will allow us
to compare the detailed model line profiles to our echelle
observations for a subset of our sample stars.

\acknowledgements  LMW thanks Geoff Marcy and Gibor Basri for
providing generous access to the California Planet Search HIRES
spectra, as well as Chris Johns-Krull and Andrew West for
illuminating conversations and assistance in data analysis. LMW
and SLH additionally thank Jeffrey Linsky for his valuable comments as
our referee, which have greatly improved this publication. Support for
this work, associated with HST program GO-10525, was provided by NASA
through a grant from the Space Telescope Science Institute, which is
operated by the Association of Universities for Research in Astronomy,
Inc., under NASA contract NAS 5-26555.  The authors also wish to
recognize and acknowledge the very significant cultural role and
reverence that the summit of Mauna Kea has always had within the
indigenous Hawaiian community.

\clearpage

\pagestyle{empty}

\begin{figure}
\begin{center}
\includegraphics[width=0.7\textwidth]{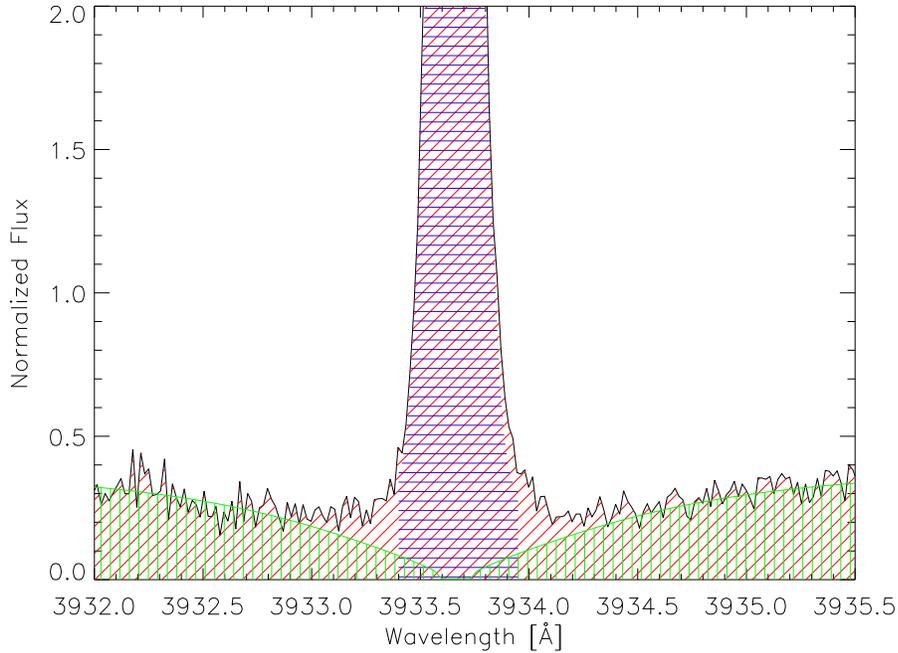}
\end{center}
\caption{The total flux in Ca II K (red diagonal hatching) is comprised of a chromospheric emission core (blue horizontal hatching) superimposed over a photospheric absorption line (green vertical hatching). However, some flux in the inner
  line wings is also chromospheric, i.e. the flux exceeds
  that predicted by the radiative equilibrium model absorption line
  (green vertical hatching). Our method of measuring the Ca II K line corrects for the photospheric contribution to the line by subtracting off the radiative equilibrium profile, thereby measuring only the chromospheric flux in Ca II K.}
\label{wingdiff}
\end{figure}

\begin{figure}[htp]
  \begin{center}
    \subfigure[]{\label{havcak_corrected}\includegraphics[width=0.45\textwidth]{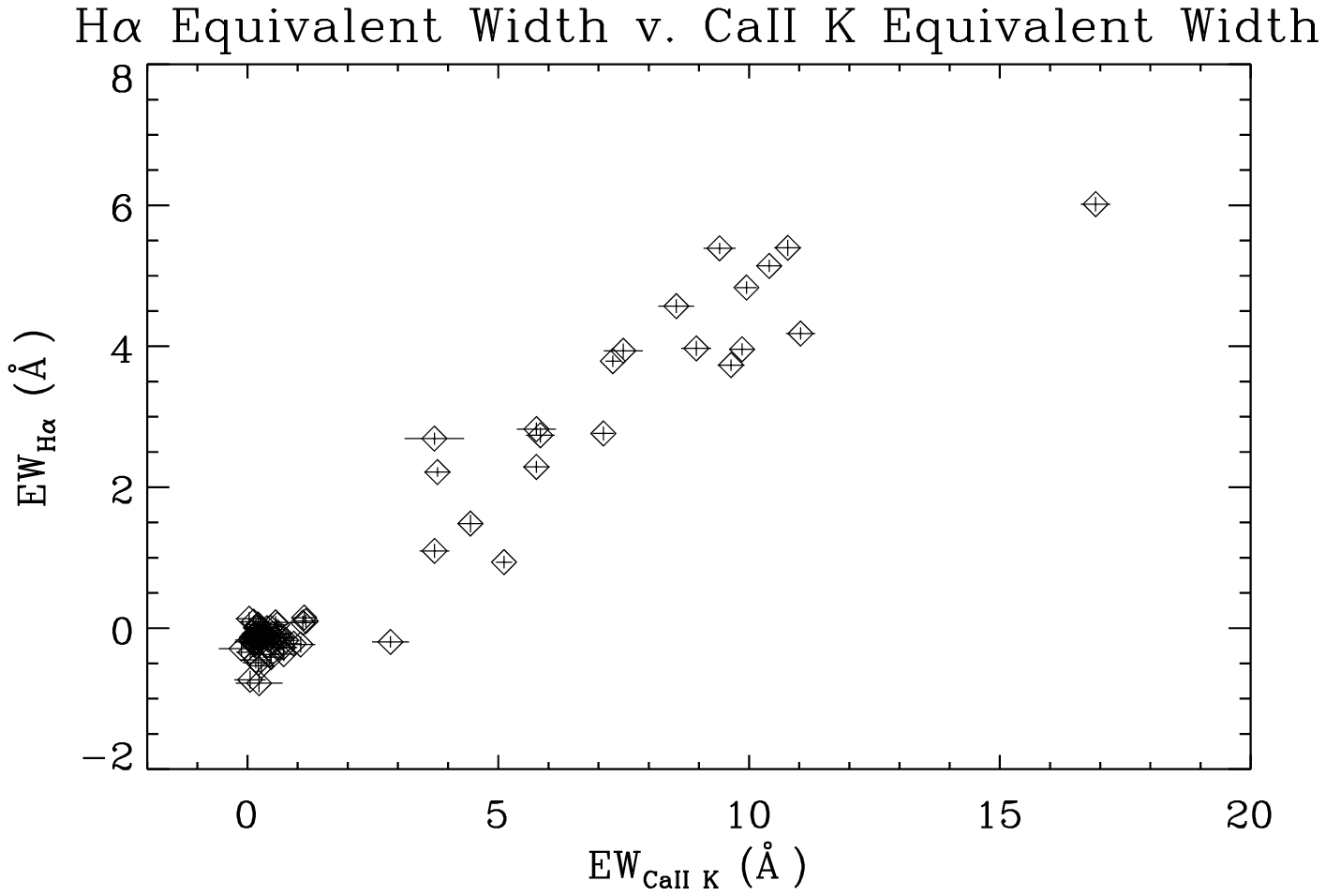}}
    \subfigure[]{\label{havcak_corrected_zoom}\includegraphics[width=0.45\textwidth]{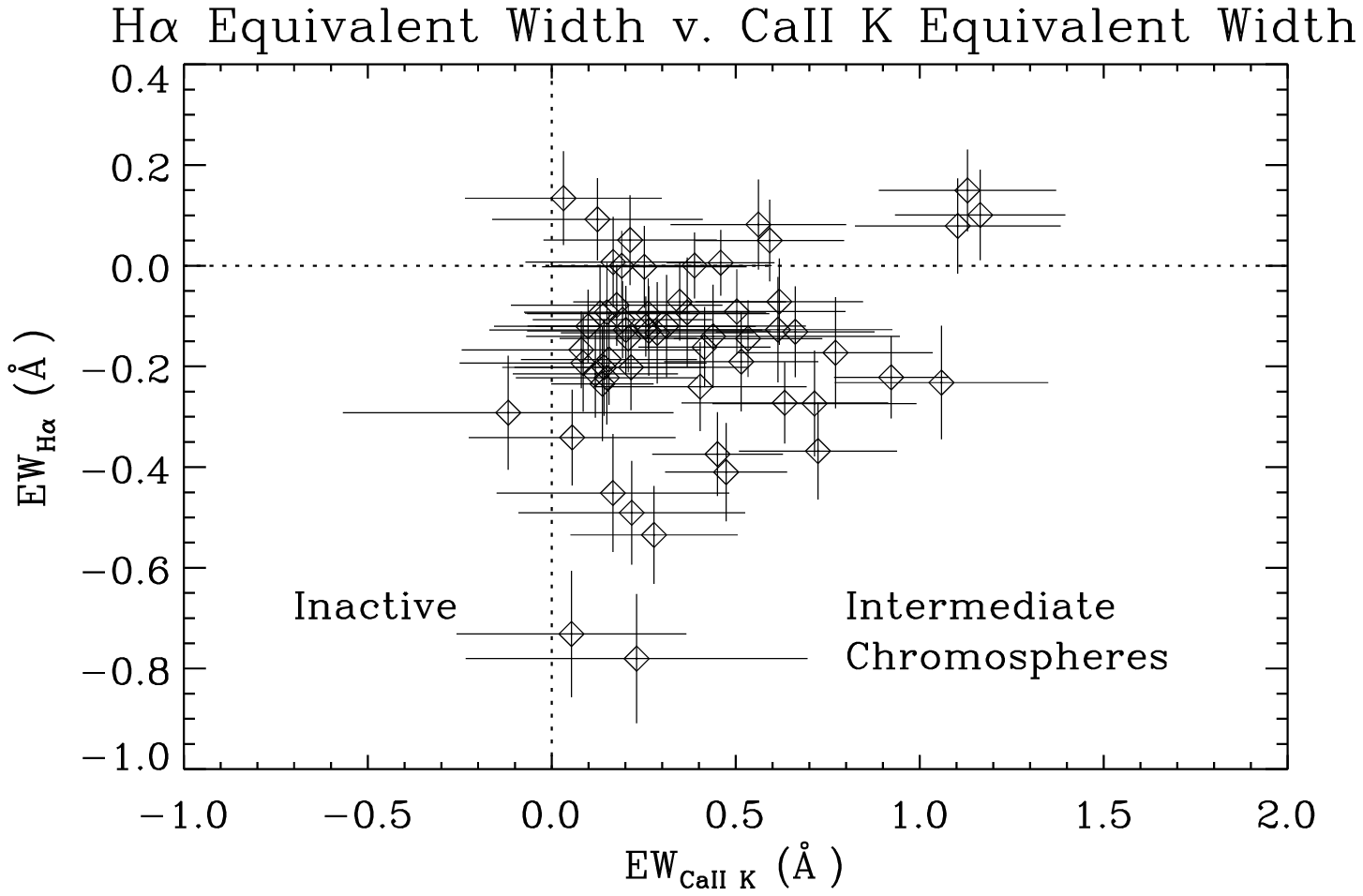}}
\end{center}
  \caption{(a): Total H$\alpha$ equivalent width versus {\em
  chromospheric} Ca II K for the entire sample observed with DIS. For low to
  intermediate activity objects (those with H$\alpha$ absorption)
  there is no obvious correlation with Ca II K. However, active stars
  show a positive correlation between the lines. (b): Total H$\alpha$
  equivalent width versus {\em chromospheric} Ca II K, shown at an
  expanded scale for the weak to intermediate activity stars. The
  perpendicular dotted lines plotted over the data indicate zero H$\alpha$
  and Ca II K equivalent widths-- stars with H$\alpha$ absorption
  (negative equivalent width) would traditionally be classified as
  ``inactive'', but many of these stars possess small chromospheric Ca II K
  emission. Increasing absorption in H$\alpha$ as Ca II K increases is not evident
  from these low resolution data.}
  \label{}
\end{figure}

\begin{figure}
  \begin{center}
    \subfigure[]{\label{havcak_corrected_echelle}\includegraphics[width=0.55\textwidth]{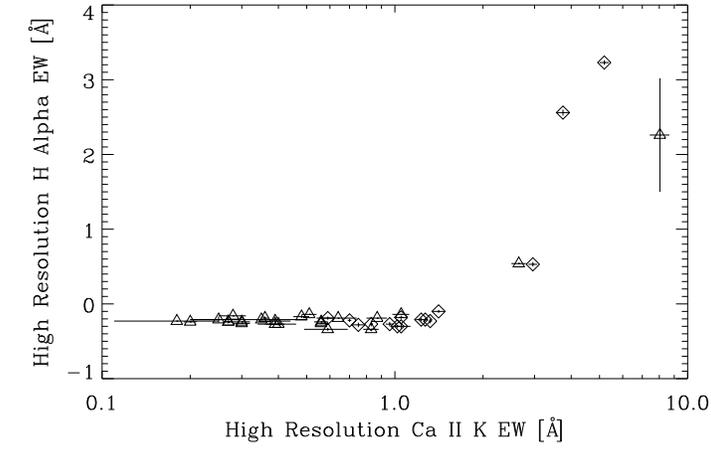}}
    \subfigure[]{\label{havcak_corrected_echelle_ZOOM}\includegraphics[width=0.55\textwidth]{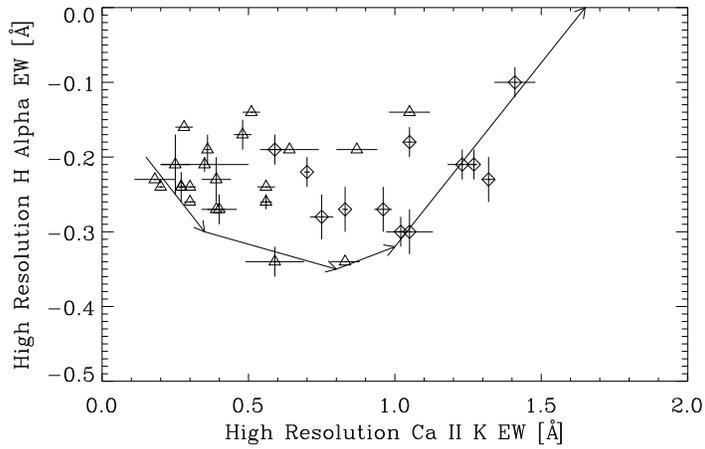}}
\end{center}
\caption{(a:) H$\alpha$ and Ca II K equivalent widths measured from
  our echelle spectra. Diamonds denote measurements of our ARCES echelle data, while triangles indicate measurements from our HIRES data. As in the case of the low resolution
  observations, H$\alpha$ absorption is associated with a range
  in Ca II K emission, while H$\alpha$ emission seems to be positively
  correlated with Ca II K. The weak to intermediate activity objects,
  plotted at an expanded scale in (b), seem to indicate an initial
  increase in H$\alpha$ absorption with increasing Ca II K, as per the
  observations of \citet{sh86}. The arrows indicate the schematic locus \citep[predicted by][]{cram87} of an initial increase in H$\alpha$ absorption as Ca II K increases, and subsequent filling in
  with emission as Ca II K increases even further. Interestingly, the locus appears to delineate the envelope of the observations.}
\label{}
\end{figure}

\begin{figure}
\begin{center}
\includegraphics[width=0.7\textwidth]{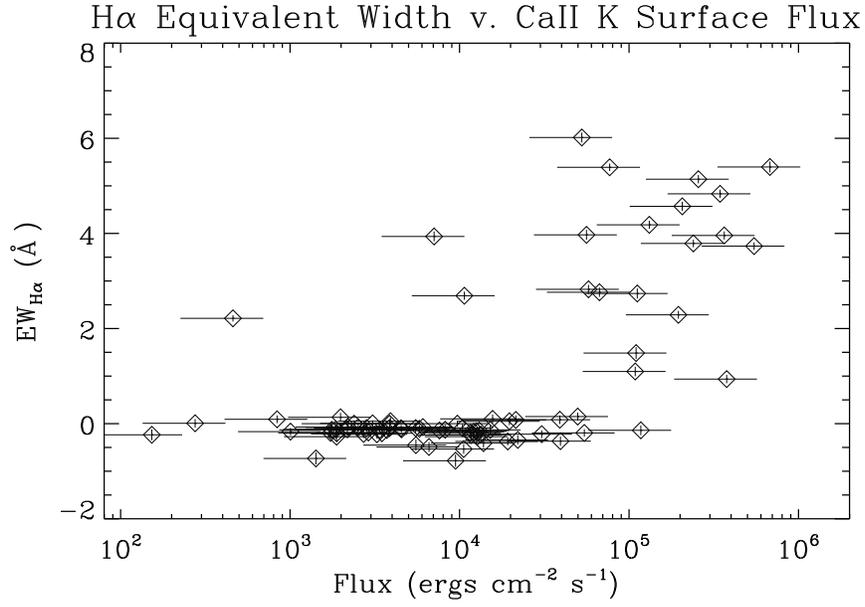}
\end{center}
\caption{H$\alpha$ equivalent width versus the Ca II K surface
  flux measured from our low resolution DIS data. More active stars (those with larger Ca II K flux) have a wide
  range of scatter in their H$\alpha$ equivalent widths.}
\label{haewvcaksurf}
\end{figure}

\begin{figure}[htp]
  \begin{center}
    \subfigure{\label{fig:edge-a}\includegraphics[width=0.4\textwidth]{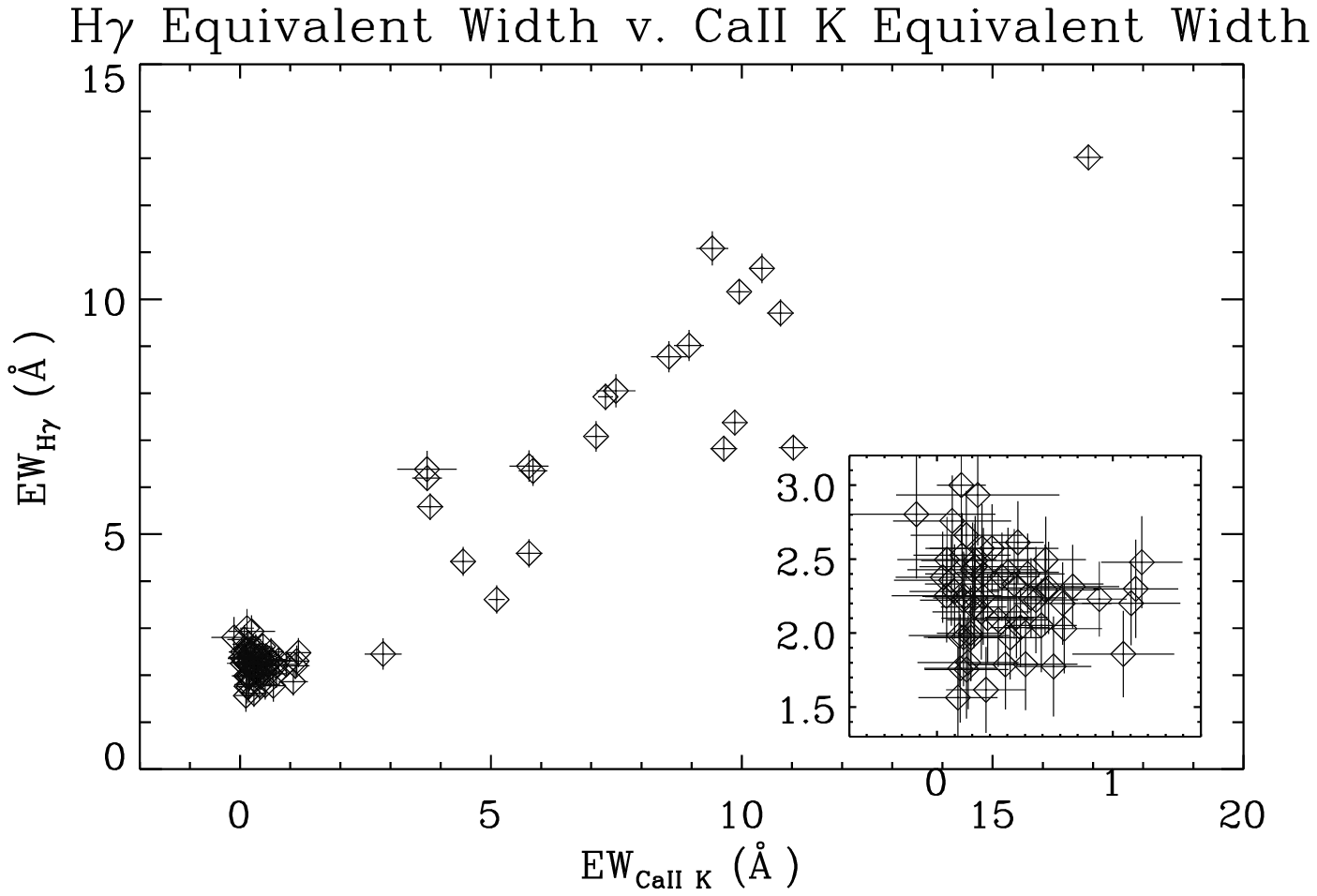}}
    \subfigure{\label{fig:edge-b}\includegraphics[width=0.4\textwidth]{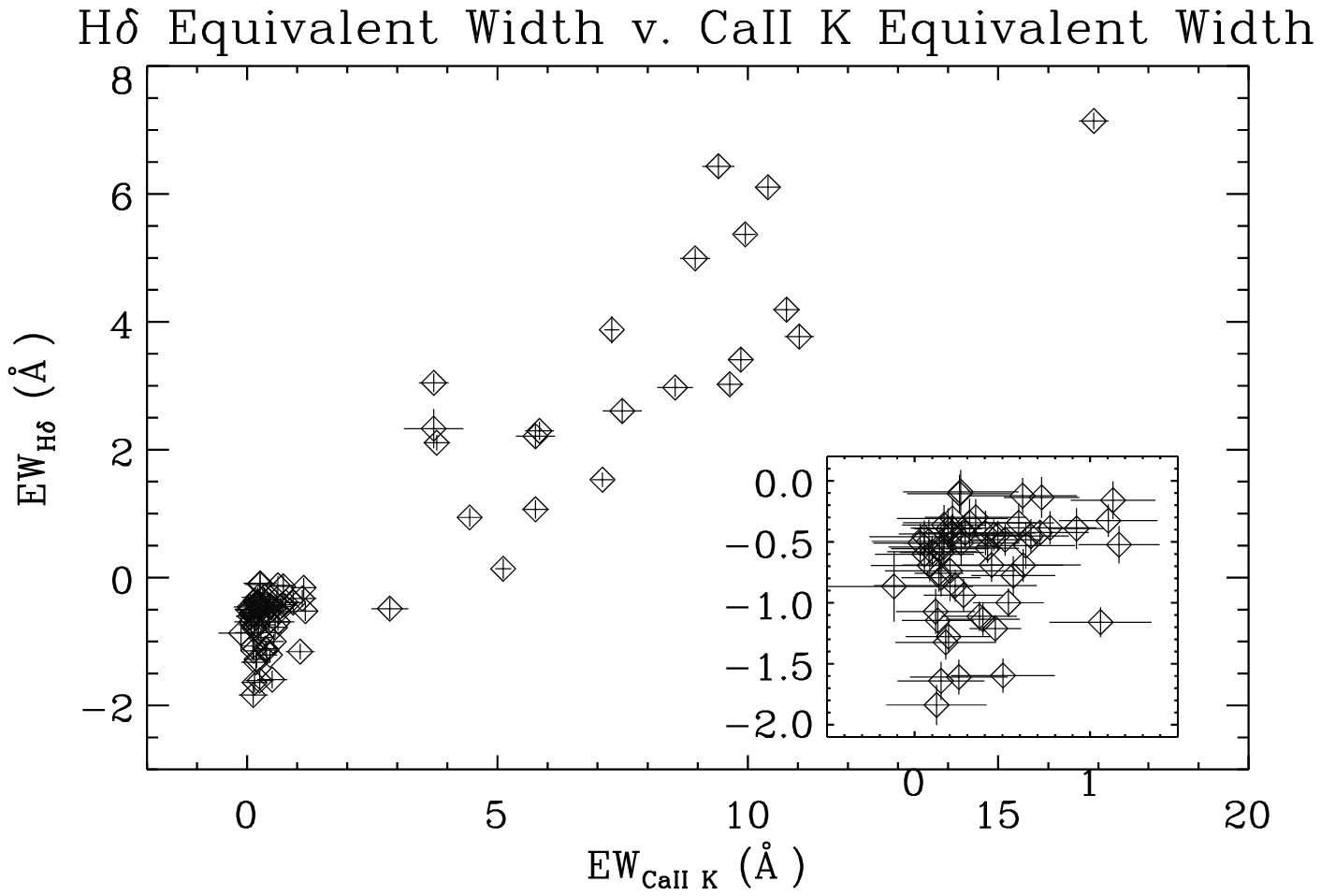}} \\
    \subfigure{\label{fig:edge-c}\includegraphics[width=0.4\textwidth]{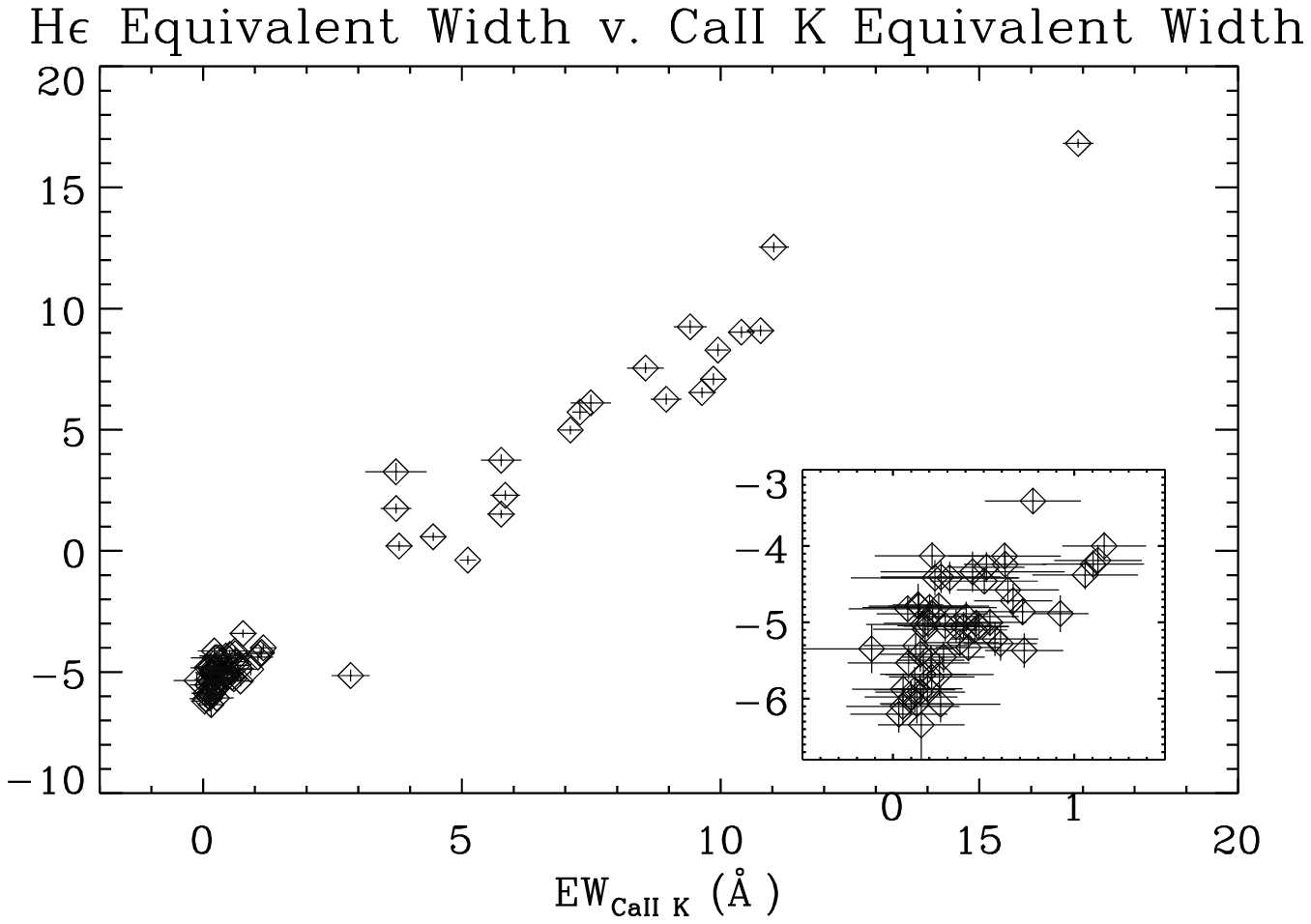}}
    \subfigure{\label{fig:edge-c}\includegraphics[width=0.4\textwidth]{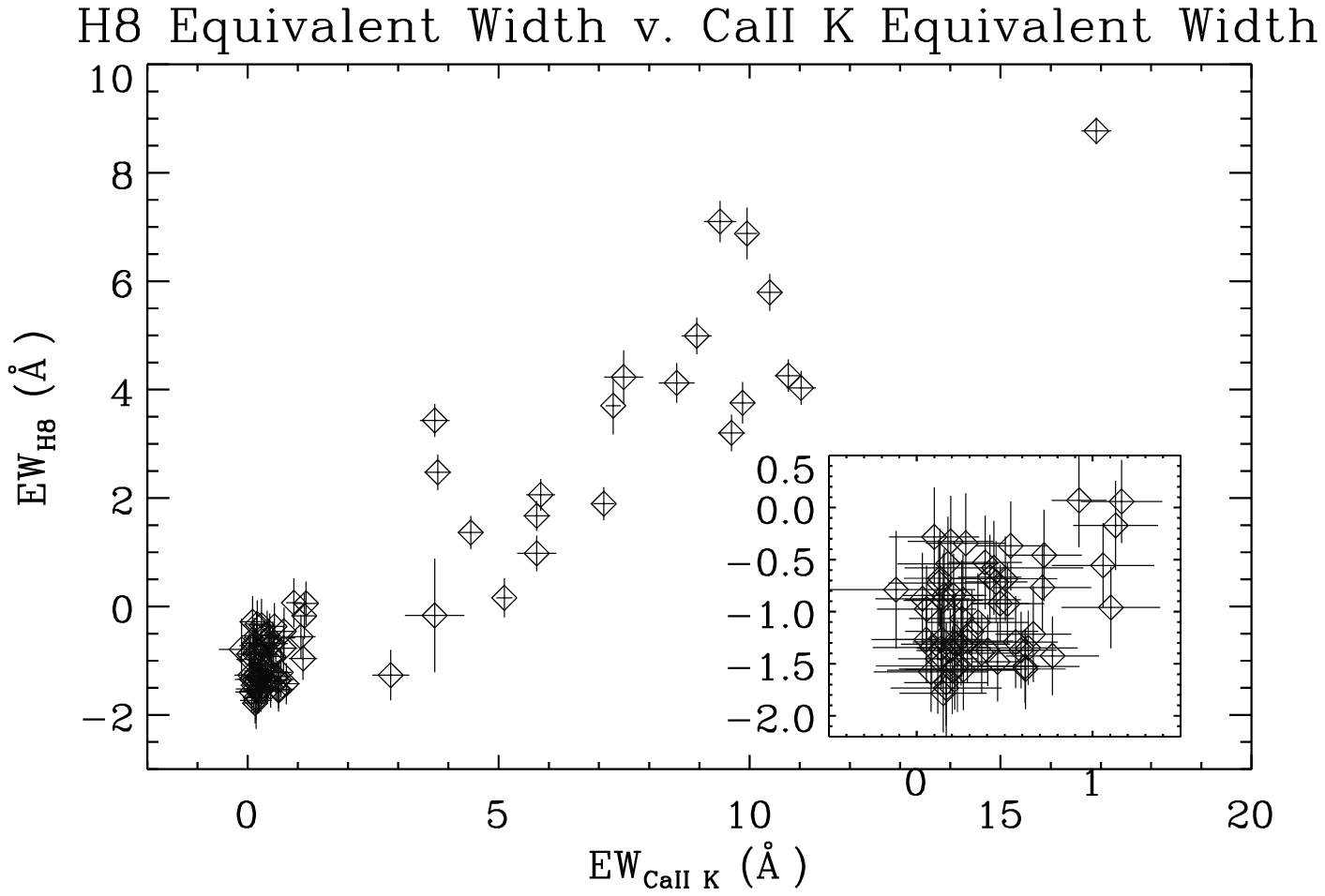}}
  \end{center}
  \caption{Equivalent widths for higher order Balmer lines versus Ca
  II K equivalent widths from DIS data. The higher order line and Ca II K  show a similar relationship as Ca II K did with H$\alpha$ in Figure 4(a).}
  \label{higherlines}
\end{figure}

\begin{figure}
\begin{center}
\includegraphics[width=0.7\textwidth]{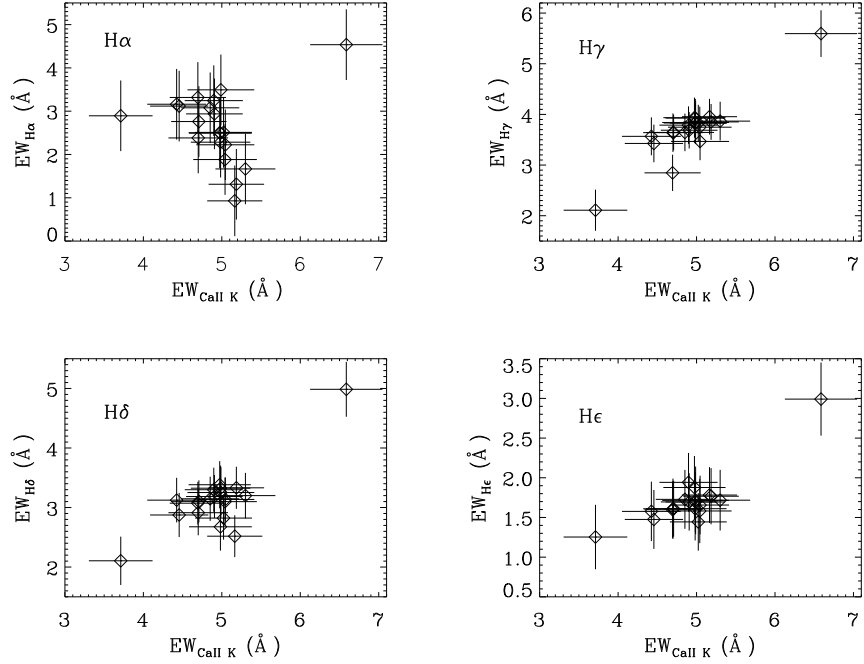}
\end{center}
\caption{Multiple measurements of equivalent widths of the Balmer lines and
  Ca II K in the active star AD Leo, measured from HIRES echelle data. Although a positive correlation
  between the instantaneous measurements of H$\alpha$ and Ca II K is evident in Figures 2 and 3, it is clear
  that the two lines are not neccessarily positively correlated in time-resolved observations of a single star. }
\label{thesis_adleo_multiplot}
\end{figure}

\begin{figure}[htp]
  \begin{center}
    \subfigure{\label{fig:edge-a}\includegraphics[width=0.4\textwidth]{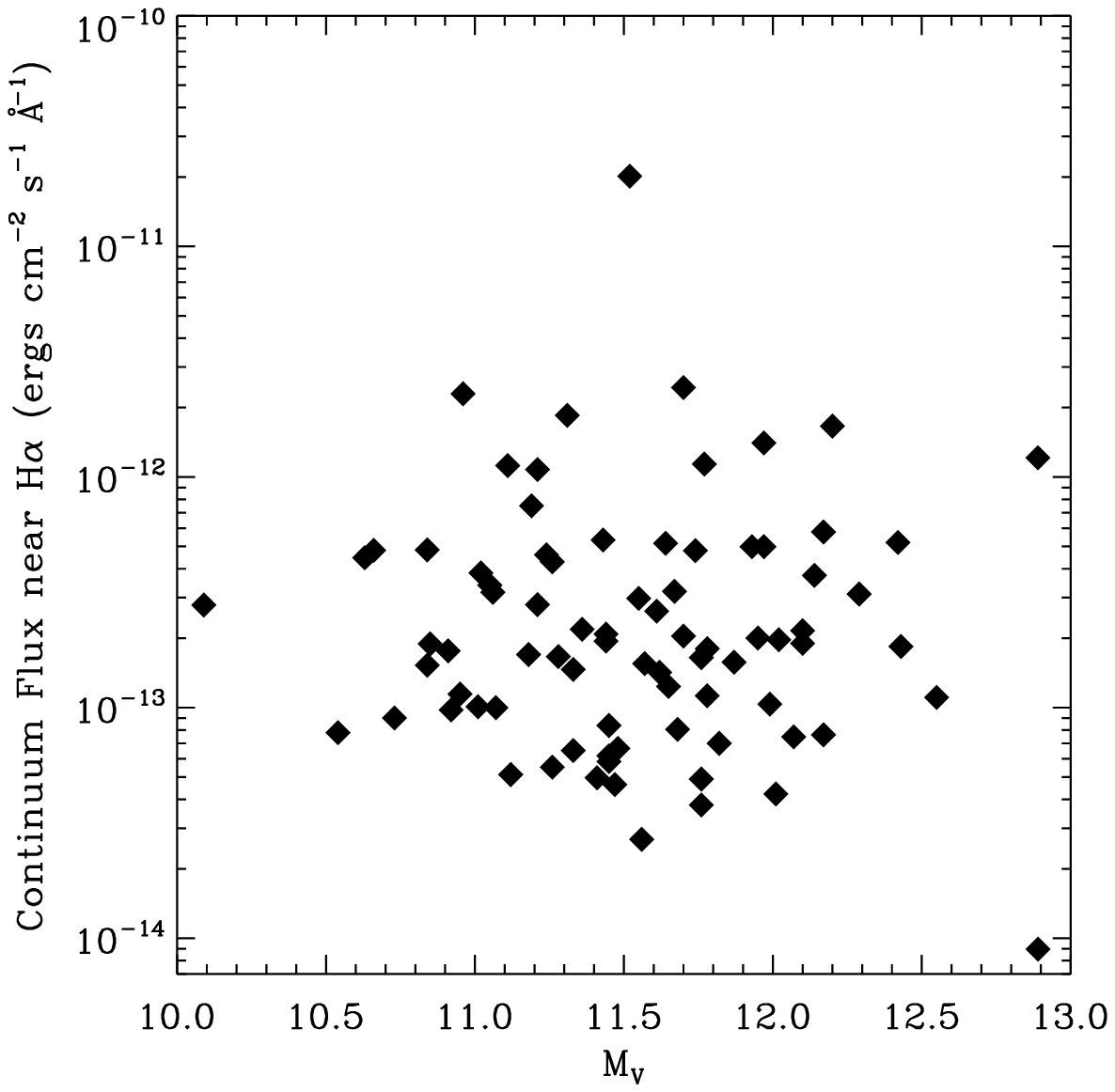}}
    \subfigure{\label{fig:edge-b}\includegraphics[width=0.4\textwidth]{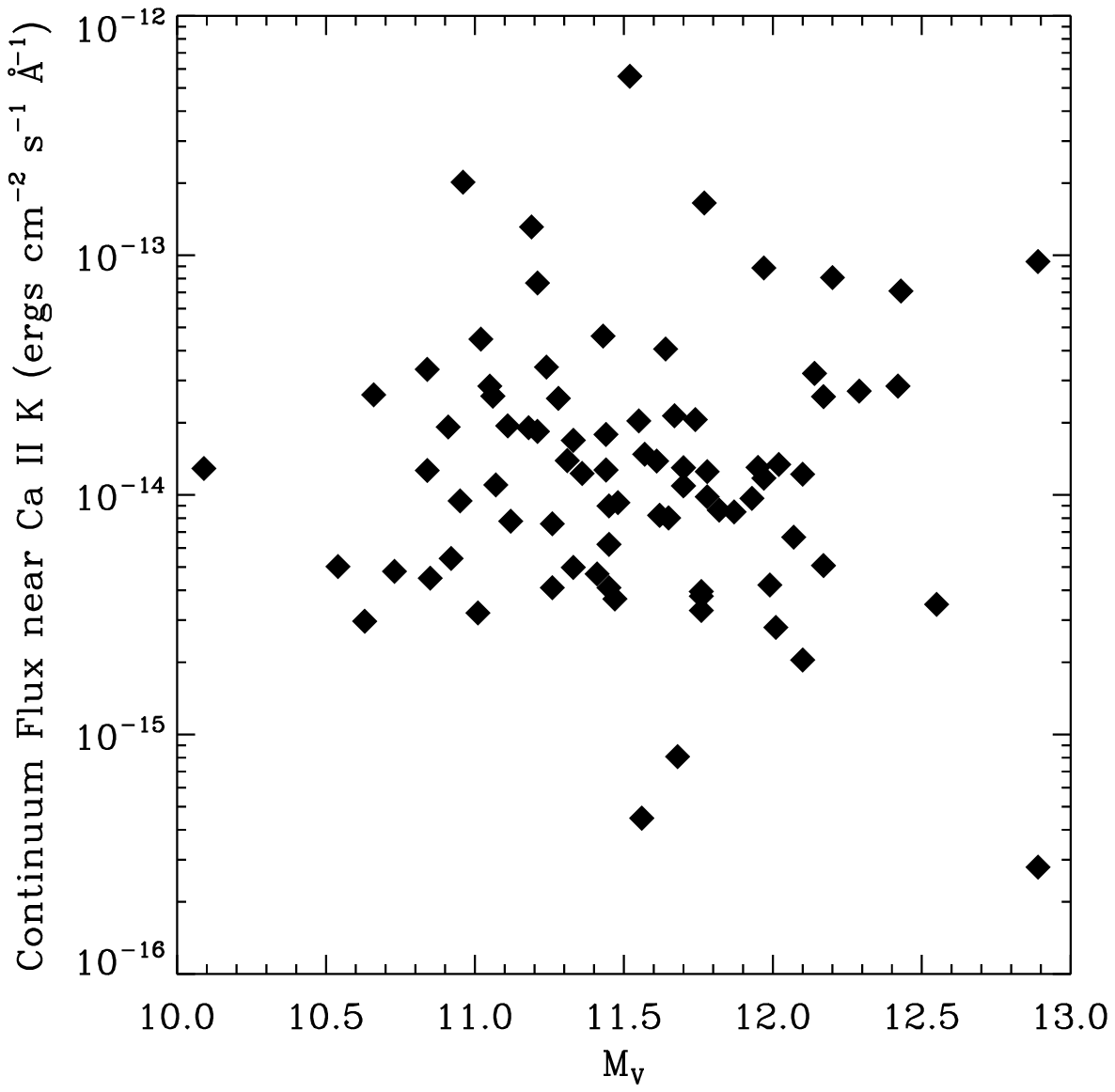}} 
   \end{center}
  \caption{Continua near H$\alpha$ and
  Ca II K from our flux-calibrated DIS data versus absolute V magnitude. There is no discernable trend between the continuum level
  and the absolute magnitude of the stars in our sample. }
  \label{mvcont}
\end{figure}

\begin{figure}[htp]
  \begin{center}
    \subfigure{\label{fig:edge-a}\includegraphics[width=0.4\textwidth]{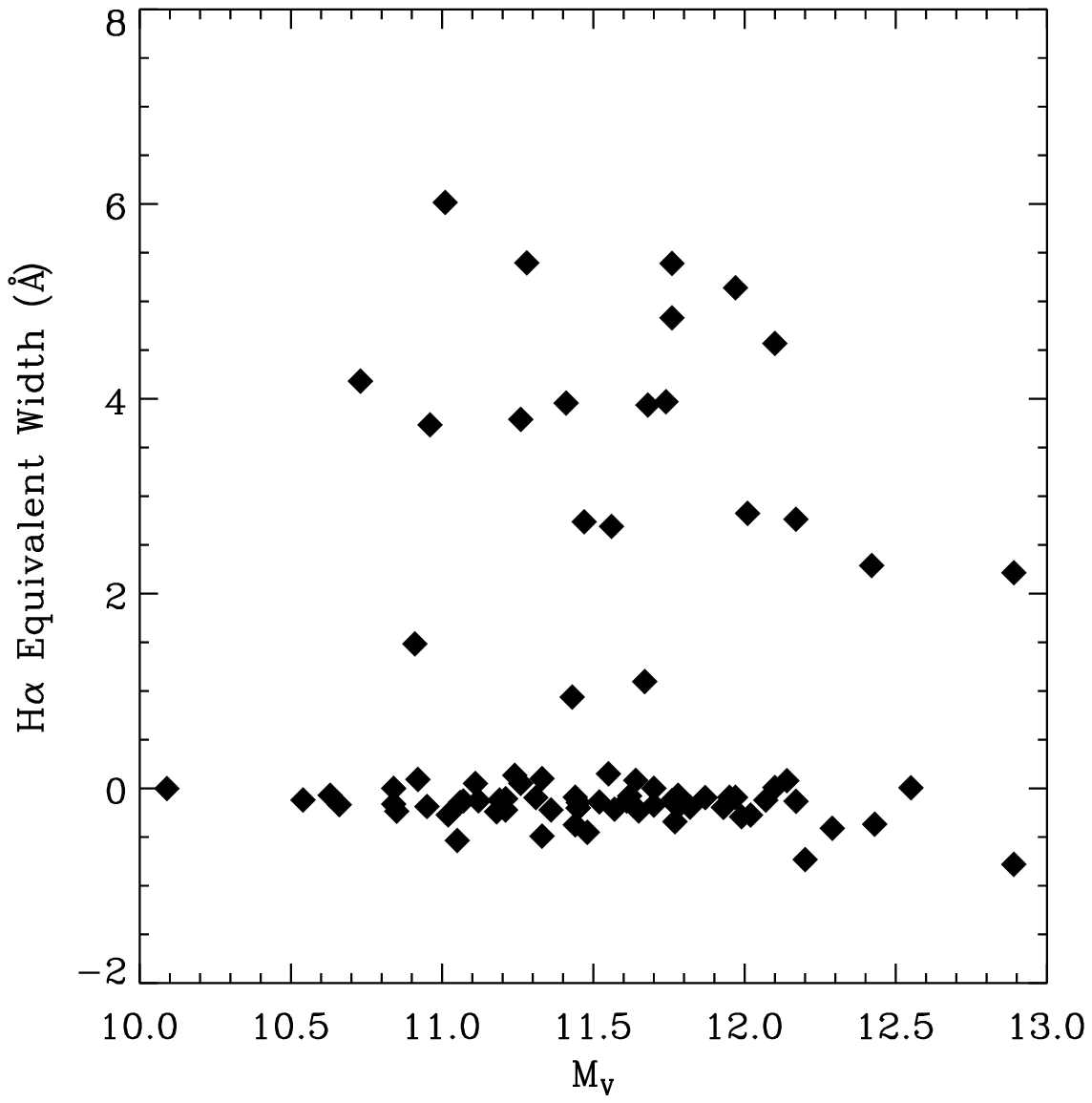}}
    \subfigure{\label{fig:edge-b}\includegraphics[width=0.4\textwidth]{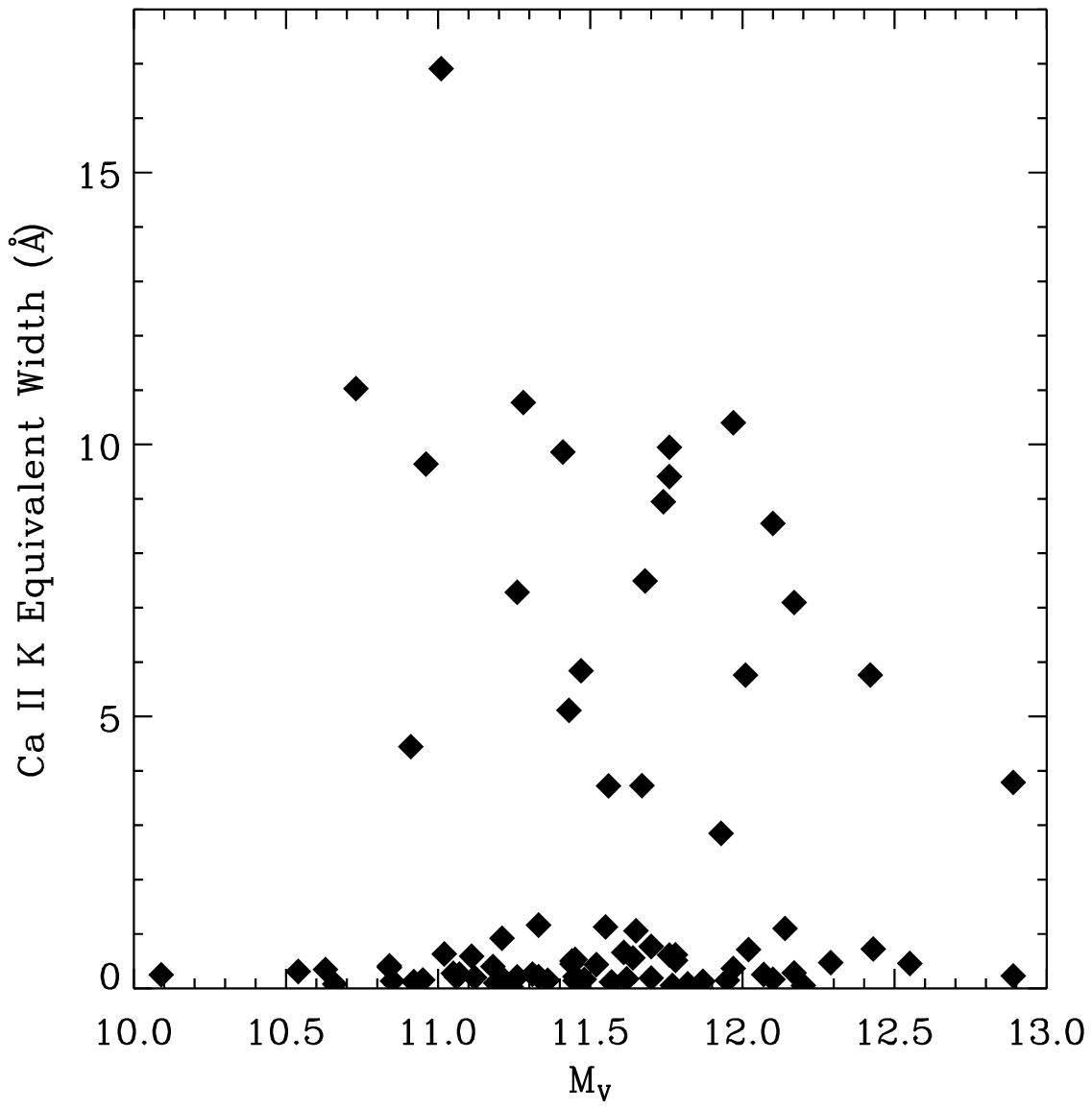}} 
   \end{center}
  \caption{Equivalent widths of H$\alpha$ and
  Ca II K (measured from DIS data) versus absolute V magnitude. The activity levels we determine for each star show no
  bias towards higher luminosity stars.}
  \label{mvlines}
\end{figure}

\begin{figure}
  \begin{center}
\includegraphics[width=0.7\textwidth]{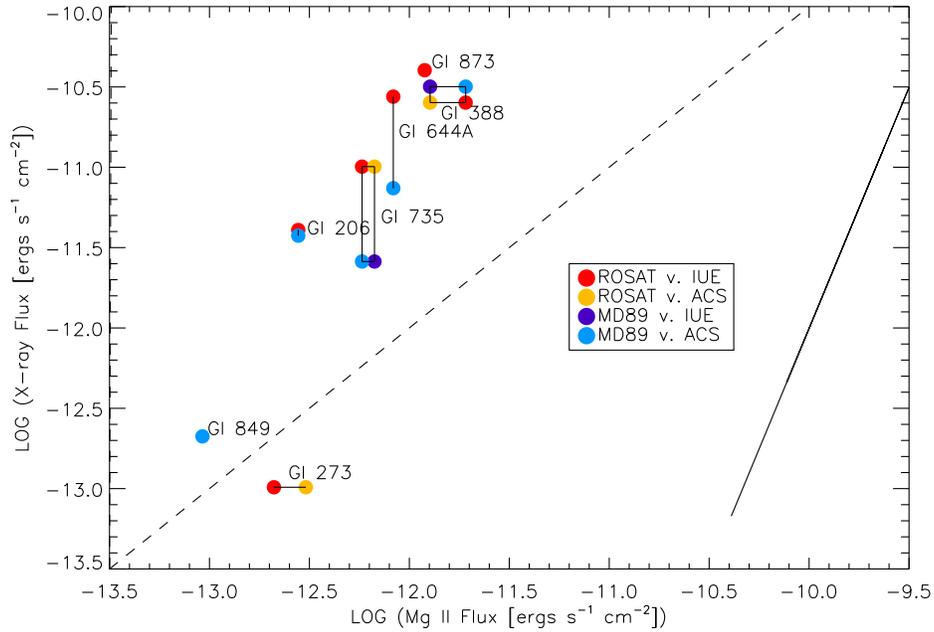}
  \end{center}
\caption{X-ray flux versus Mg II flux for a subset of our
  sample. Different plotting symbols are used to indicate measurements
  by different instruments. Points corresponding to multiple observations of a
  given star by different instruments are connected by solid lines.The
  solid line at right represents the approximate power law fit to Mg
  II and soft X-ray emission of G
  and K dwarfs given in \citet{ayres81}, while the dashed line
  indicates the line of equality between the Mg II and X-ray
  emission. }
\label{xrayvmg}
\end{figure}

\begin{figure}
  \begin{center}
    \subfigure[X-ray versus H$\alpha$ flux]{\label{}\includegraphics[width=0.45\textwidth]{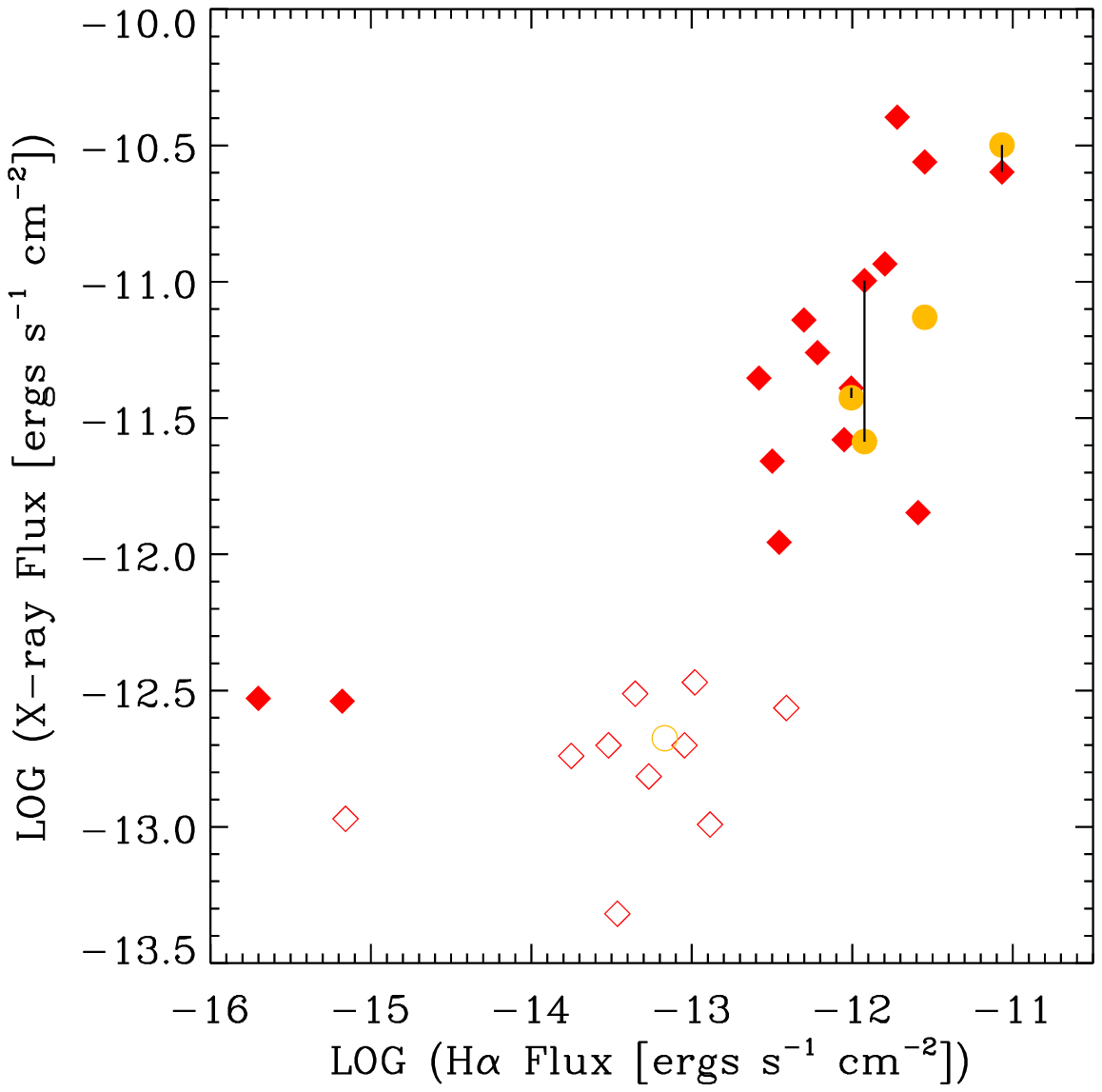}}
    \subfigure[X-ray versus Ca II K flux]{\label{}\includegraphics[width=0.45\textwidth]{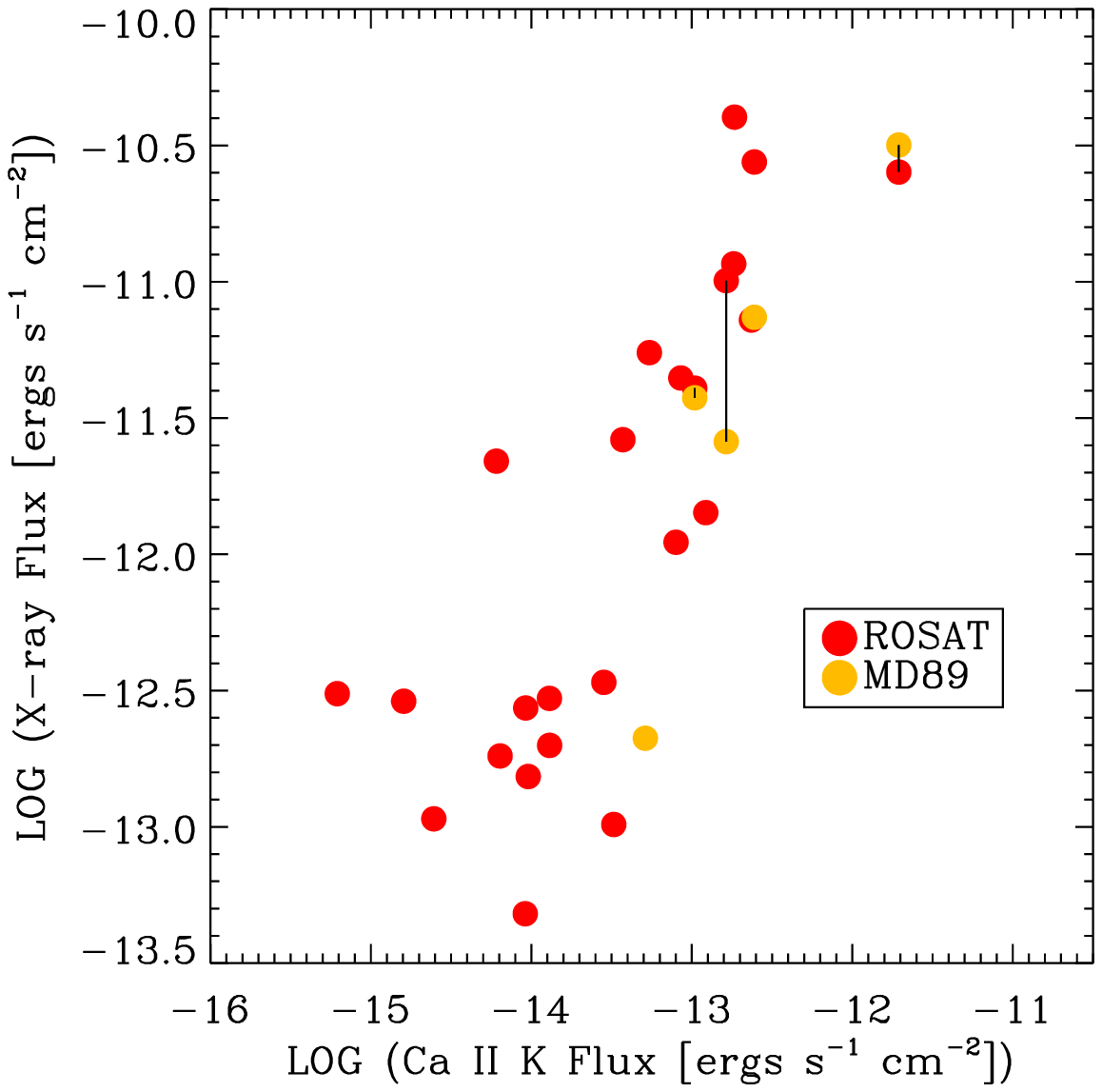}}
  \end{center}
\caption{X-ray flux versus the optical chromospheric flux as measured
  by H$\alpha$ and Ca II K (measured from flux-calibrated DIS data). In (a), open circles correspond to H$\alpha$ absorption, while filled circles denote H$\alpha$ emission. As the Ca II K line flux we report is always due to the chromospheric emission core, all symbols in (b) are filled to indicate emission. The three points that lie to the left of
  the main locus in (a) are H$\alpha$ lines with very low equivalent
  width, probably due to their being almost completely filled in by
  chromospheric emission. }
\label{xrayvopt}
\end{figure}

\begin{figure}
  \begin{center}
    \subfigure[Mg II versus H$\alpha$ flux]{\label{}\includegraphics[width=0.45\textwidth]{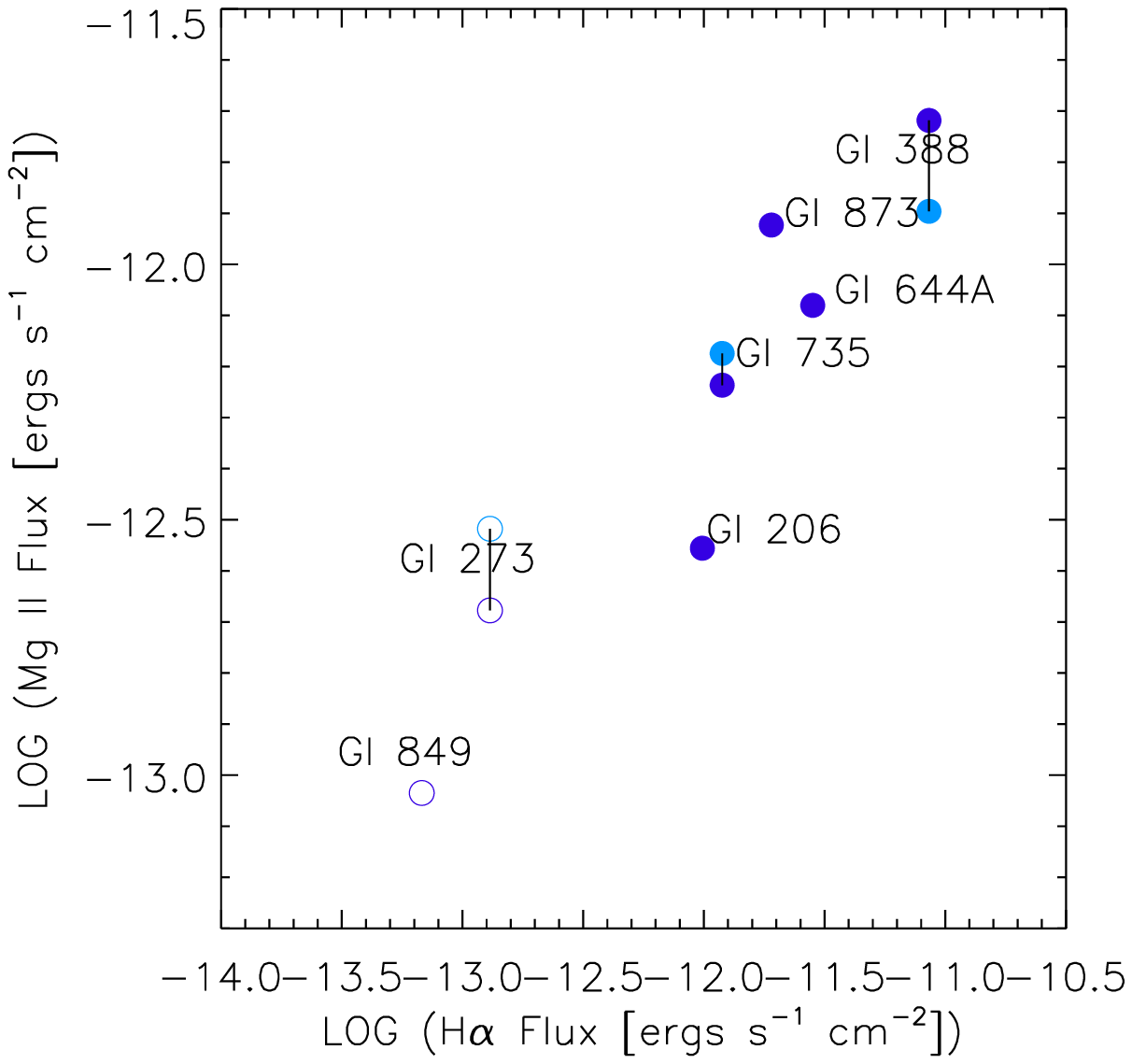}}
    \subfigure[Mg II versus Ca II K flux]{\label{}\includegraphics[width=0.45\textwidth]{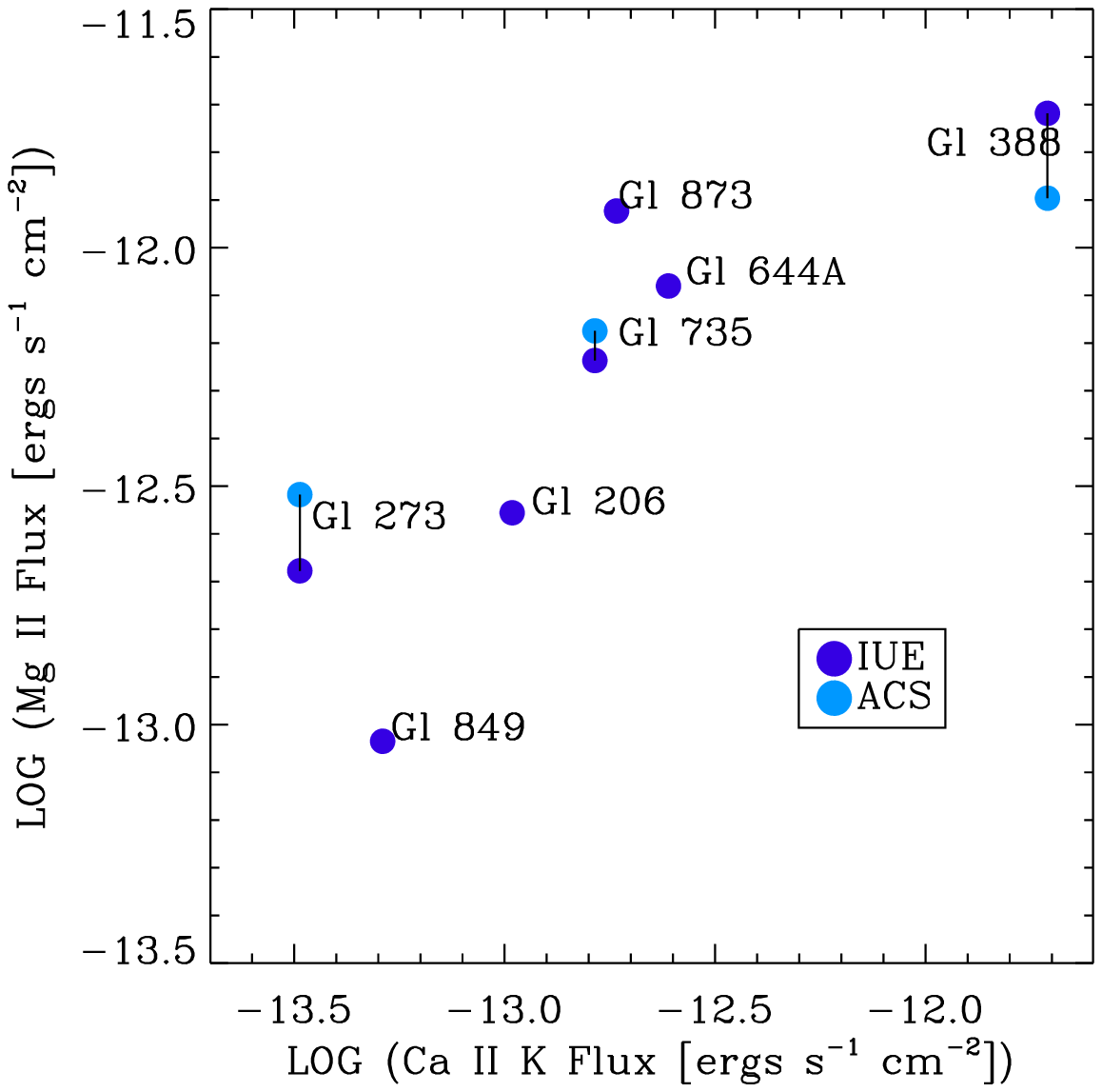}}
  \end{center}
\caption{UV chromospheric (Mg II) flux versus the optical chromospheric flux as measured
  by H$\alpha$ and Ca II K. Different plotting symbols indicate Mg II
  measurements taken with either IUE or the ACS near-UV
  prism; multiple observations of stars are connected with
  solid lines. In (a), two of the stars with Mg II measurements have H$\alpha$ absorption-- these are indicated by open plotting symbols, while emission is denoted by filled symbols.  In the Mg II sample of active stars, the positive correlation is reminiscent of that for Ca II K and H$\alpha$ in our optical measurements.}
\label{mgvopt}
\end{figure}

\begin{figure}
  \begin{center}
    \subfigure[Hardness ratio versus H$\alpha$]{\label{}\includegraphics[width=0.45\textwidth]{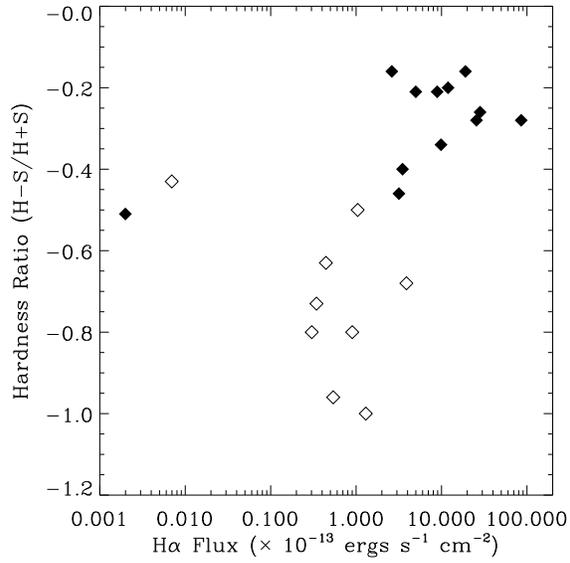}}
    \subfigure[Hardness ratio versus Ca II K]{\label{}\includegraphics[width=0.45\textwidth]{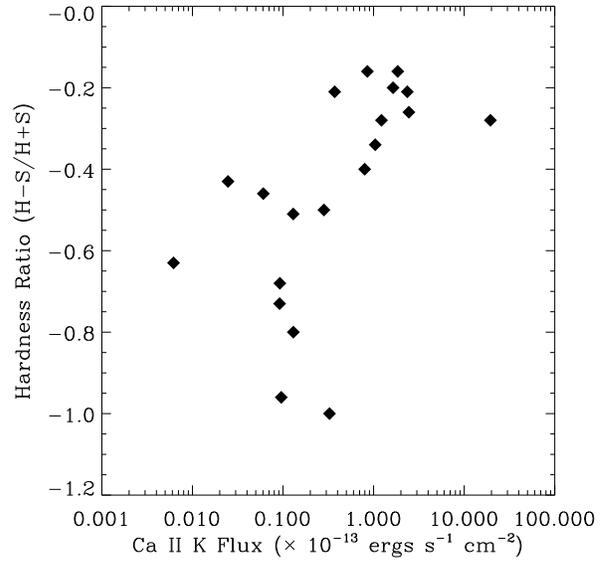}}
  \end{center}
\caption{Hardness ratio versus measurements of H$\alpha$ and Ca II K
  flux from DIS data. Open diamonds in (a) denote H$\alpha$ absorption, while filled diamonds indicate emission. Increasing hardness is indicative of hotter
  coronal temperatures, and correlates with increasing
  flux in the chromospheric lines.}
\label{hardnessvopt}
\end{figure}

\begin{figure}
  \begin{center}
\includegraphics[width=0.7\textwidth]{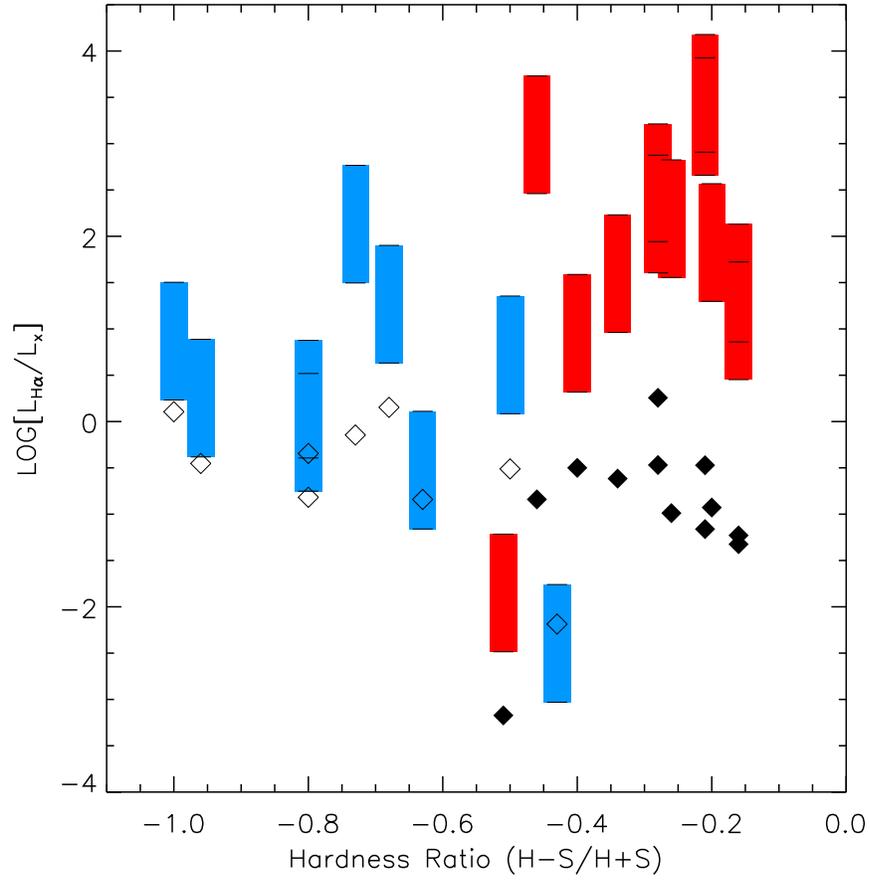}
  \end{center}
\caption{Ratio of the luminosity in H$\alpha$ to the X-ray luminosity
  versus the hardness ratio. Open diamonds indicate H$\alpha$ absorption, while filled symbols indicate H$\alpha$ emission. Stars lie at a roughly constant value of
  L$_{H\alpha}$/L$_x$ $\sim$ -0.6, consistent with observations of
  both field and cluster M dwarfs.}
\label{hardness}
\end{figure}

\clearpage  

\begin{deluxetable}{llrrrrl} 
\tablecolumns{7} 
\tablewidth{0pc} 
\tablecaption{Summary of Program Stars} 
\tablehead{\colhead{CNS3} & \colhead{Name} & \colhead{$\alpha$ 2000} &
  \colhead{$\delta$ 2000} &  \colhead{M$_V$} &  \colhead{d (pc)} &  \colhead{SpT}}
\tablerefs{\citet{re95a}}
\startdata 
460 &  GJ 3168  &  02 36 27.2  &  $+$55 28 35  & 11.45 & 22.6 &  M 3 \\ 
492 &  Gl 109  &  02 44 14.9  &  $+$25 31 25  & 11.07 & 8 &  M 3 \\ 
836 &  Gl 179  &  04 52 05.6  &  $+$06 28 37  & 11.65 & 11.6 &  M 3.5 \\ 
876 &  GJ 3333  &  05 07 49.2  &  $+$17 58 59  & 11.24 & 12.8 &  M 3   \\ 
919 &  Gl 203  &  05 28 00.1  &  $+$09 38 43  & 12.42 & 10.3 &  M 3.5 \\ 
933 &  Gl 206  &  05 32 14.6  &  $+$09 49 15  & 10.95 & 13 &  M 3.5 \\ 
941 &  GJ 3356  &  05 34 52.1  &  $+$13 52 48  & 11.44 & 11.8 &  M 3.5 \\ 
1068 &  GJ 3404A  &  06 42  11.1  &  $+$03 34 54  & 11.62 & 12.2 &  M 3.5 \\ 
1089 &  GJ 3412  &  06 54 03.7  &  $+$60 52 24  & 10.91 & 10.7 &  M 3   \\ 
1094 &  Gl 251  &  06 54 49.4  &  $+$33 16 07  & 11.21 & 5.8 &  M 3   \\ 
1126 &  Gl 263  &  07 04 17.2  &  $-$10 30 08  & 10.63 & 13.5 &  M 3.5 \\ 
1168 &  Gl 273  &  07 27 30.8  &  $+$05 13 12  & 11.97 & 3.8 &  M 3.5 \\ 
1184 &  Gl 277B  &  07 31 57.4  &  $+$36 13 48  & 11.43 & 11.7 &  M 3.5 \\ 
1226 &  GJ 1101  &  07 55 55.5  &  $+$83 23 08  & 12.55 & 12.8 &  M 3.5 \\ 
1247 &  GJ 1105  &  07 58 12.5  &  $+$41 18 19  & 12.43 & 8.2 &  M 3.5 \\ 
1328 &  GJ 2069A  &  08 31 37.6  &  $+$19 23 39  & 12.1 & 9.1 &  M 3.5 \\ 
1353 &  GJ 3511B  &  08 39 48.2  &  $+$08 56 16  & 12.07 & 15.1 &  M 3.5 \\ 
1405 &  GJ 3522  &  08 58  56.1  &  $+$08 28 28  & 12.17 & 5.6 &  M 3.5 \\ 
1420 &  GJ 3527  &  09 01  10.7  &  $+$01 56 35  & 11.36 & 12.3 &  M 3   \\ 
1441 &  GJ 3537  &  09 09 39.2  &  $+$06 42 08  & 11.26 & 26.2 &  M 3  \\ 
1501 &  GJ 1125  &  09 30 44.8  &  $+$00 19 25  & 11.68 & 10.1 &  M 3.5 \\ 
1507 &  Gl 352A  &  09 31 18.9  &  $-$13 29 18  & 11.01 & 9.1 &  M 3   \\ 
1530 &  Gl 363  &  09 42 23.8  &  $+$55 59 05  & 11.78 & 14 &  M 3.5 \\ 
1529 &  Gl 362  &  09 42 52.5  &  $+$70 02 23  & 11.05 & 10.8 &  M 3   \\ 
1611 &  Gl 386  &  10 16 46.1  &  $-$11 57 36  & 10.92 & 10.3 &  M 3   \\ 
1616 &  Gl 388     &  10 19  36.4  &  $+$19 52 11  & 10.96 & 4.9 &  M 3   \\ 
1657 &  GJ 361 &  10 35 29.0  &  $+$69 27 03  & 11.45 & 12.5 &  M 3.5 \\ 
1658 &  Gl 398  &  10 36 01.5  &  $+$05 07 10  & 11.97 & 13.4 &  M 3.5 \\ 
1660 &  GJ 3613  &  10 38 27.4  &  $+$48 32 05  & 11.41 & 26 &  M 3  \\ 
1690 &  Gl 403  &  10 52 04.6  &  $+$13 59 48  & 12.1 & 13 &  M 3.5 \\ 
1742 &  GJ 3647  &  11 11  51.8  &  $+$33 32 09  & 11.76 & 13.3 &  M 3.5 \\ 
1795 &  GJ 3666  &  11 28 56.5  &  $+$10 10 34  & 11.77 & 14.4 &  M 3.5 \\ 
1838 &  Gl 443  &  11 46 42.5  &  $-$14 00 43  & 11.21 & 12.4 &  M 3   \\ 
1845 &  Gl 445  &  11 47 39.2  &  $+$78 41 24  & 12.2 & 5.3 &  M 3.5 \\ 
1931 &  GJ 3719  &  12 16 58.5  &  $+$31 09 22  & 11.76 & 30 &  M 3.5 \\ 
1988 &  Gl 480  &  12 38 53.1  &  $+$11 41 47  & 11.11 & 12.1 &  M 3  \\ 
2009 &  Gl 486  &  12 47 57.2  &  $+$09 45 09  & 11.78 & 8.3 &  M 3.5 \\ 
2088 &  GJ 3775  &  13 18  01.9  &  $+$02 14 00  & 11.76 & 17.4 &  M 3.5 \\ 
2094 &  Gl 507B  &  13 19 34.4  &  $+$35 06 30  & 11.45 & 13.5 &  M 3  \\ 
2175 &  GJ 3804  &  13 45 50.9  &  $-$17 57 56  & 11.67 & 10.9 &  M 3.5 \\ 
2300 &  Gl 553.1  &  14 31 01.4  &  $-$12 17 43  & 11.7 & 11.1 &  M 3.5 \\ 
2399 &  GJ 3892  &  15 09 35.9  &  $+$03 09 56  & 11.18 & 11.4 &  M 3  \\ 
2420 &  Gl 581  &  15 19 27.5  &  $-$07 43 20  & 11.52 & 6.4 &  M 3   \\ 
2456 &  GJ 3913  &  15 35 46.5  &  $+$22 09 46  & 11.48 & 17.5 &  M 3.5 \\ 
2472 &  Gl 597  &  15 41 14.7  &  $+$75 59 38  & 11.55 & 13.6 &  M 3   \\ 
2567 &  Gl 617B  &  16 16 45.8  &  $+$67 15 20  & 10.54 & 10.9 &  M 3   \\ 
2587 &  GJ 3953  &  16 25 32.4  &  $+$26 01 37  & 11.33 & 15.2 &  M 3  \\ 
2599 &  Gl 628  &  16 30 18.0  &  $-$12 39 35  & 12.02 & 4.1 &  M  3.5 \\ 
2660 &  Gl 643  &  16 55 24.6  &  $-$08 19 27  & 12.89 & 6.1 &  M 3.5 \\ 
2660 &  Gl 643  &  16 55 24.6  &  $-$08 19 27  & 12.89 & 6.1 &  M 3.5 \\ 
2661 &  Gl 644A  &  16 55 28.1  &  $-$08 20 16  & 10.66 & 6.4 &  M 3  \\
2665 &  GJ 1207  &  16 57 05.4  &  $-$04 20 52  & 12.29 & 10 &  M 3.5 \\
2700 &  Gl 655  &  17 07 07.6  &  $+$21 33 14  & 10.84 & 14.3 &  M 3   \\ 
2714 &  GJ 3992  &  17 11 34.5  &  $+$38 26 33  & 11.7 & 9.6 &  M 3.5 \\ 
2712 &  Gl 660A  &  17 11 52.4  &  $-$01 51 02  & 11.64 & 12.1 &  M 3   \\ 
2717 &  Gl 661A  &  17 12 07.5  &  $+$45 40 09  & 11.02 & 6.1 &  M 3.5\\ 
2719 &  GJ 1212  &  17 13 40.6  &  $-$08 25 11  & 11.26 & 14.4 &  M 3.5 \\ 
2744 &  Gl 669A  &  17 19 54.5  &  $+$26 30 01  & 11.31 & 10.5 &  M 3.5\\ 
2797 &  Gl 687  &  17 36 26.3  &  $+$68 20 30  & 10.85 & 4.6 &  M 3  \\ 
2846 &  GJ 4038  &  17 57  03.6  &  $+$15 46 45  & 11.44 & 14.3 &  M 3   \\ 
2851 &  GJ 4040  &  17 57 50.9  &  $+$46 35 14  & 11.12 & 13.6 &  M 3  \\ 
2888 &  GJ 4048  &  18 18 04.1  &  $+$38 46 40  & 11.57 & 11.5 &  M 3   \\ 
2895 &  GJ 1226  &  18 20 57.4  &  $-$01 02 48  & 11.87 & 14.7 &  M 3.5 \\ 
2898 &  Gl 712  &  18 22 07.3  &  $+$06 20 36  & 11.82 & 14.2 &  M 3.5 \\ 
2901 &  GJ 4055  &  18 23 28.3  &  $+$28 10 05  & 11.93 & 13 &  M 3.5 \\ 
2916 &  GJ 4062  &  18 31  58.4  &  $+$40 41 06  & 11.61 & 11.9 &  M 3.5 \\ 
2921 &  GJ 4063  &  18 34 36.5  &  $+$40 07 27  & 12.14 & 7.2 &  M 3.5 \\ 
2924 &  Gl 720B  &  18 35 26.9  &  $+$45 45 38  & 12.17 & 14.8 &  M 3.5 \\ 
2937 &  GJ 4068  &  18 35 51.8  &  $+$80 05 37  & 12.01 & 17.5 &  M 3.5 \\ 
2940 &  GJ 4070  &  18 41  59.1  &  $+$31 49 49  & 11.06 & 11 &  M 3   \\ 
2945 &  Gl 725A  &  18 42 48.0  &  $+$59 37 29  & 11.19 & 3.5 &  M 3  \\ 
2946 &  Gl 725B  &  18 42 48.0  &  $+$59 37 29  & 11.99 & 3.5 &  M 3.5 \\ 
2964 &  GJ 4083  &  18 50  45.0  &  $+$47 58 19  & 11.95 & 13 &  M 3.5 \\ 
2976 &  Gl 735  &  18 55 27.3  &  $+$08 24 09  & 10.09 & 10.1 &  M 3  \\ 
3013 &  GJ 4098  &  19 08  29.9  &  $+$32 16 53  & 11.33 & 12.4 &  M 3   \\ 
3032 &  GJ 9652A  &  19 14 39.5  &  $+$19 18 59  & 10.73 & 14.6 &  M 3.5 \\ 
3058 &  GJ 4110  &  19 26 49.3  &  $+$16 43 03  & 11.47 & 21.1 &  M 3   \\ 
3306 &  GJ 9721B  &  21 08 44.8  &  $-$04 25 17  & 11.56 & 23.4 &  M 3   \\ 
3425 &  GJ 4231  &  21 52 10.3  &  $+$05 37 35  & 11.28 & 14.6 &  M 3   \\ 
3478 &  Gl 849  &  22 09 45.2  &  $-$04 38 11  & 10.84 & 8.1 &  M 3.5 \\ 
3584 &  Gl 873  &  22 46 50.1  &  $+$44 20 05  & 11.74 & 5.1 &  M 3.5 \\ 
\enddata 
\end{deluxetable} 

\clearpage

\begin{deluxetable}{lccc} 
\tablecolumns{4}  
\tablewidth{0pc}  
\tablecaption{Wavelength Regions Used in Line Calculations}  
\tablehead{\colhead{Line} &
\colhead{Blue Continuum (\AA)} & \colhead{Line (\AA)} & \colhead{Red Continuum (\AA)}}
\startdata   

H8 & 3850.00 $-$ 3880.00 & 3884.15 $-$ 3898.15 & 3910.00 $-$ 3930.00\\  
Ca II K & 3952.80 $-$ 3956.00 &
variable\tablenotemark{1}\tablenotetext{1}{Integration region for the
  Ca II K line was chosen interactively for individual spectra.} & 3974.80 $-$ 3976.00\\   
H $\delta$ & 4060.00 $-$ 4080.00 & 4097.00 $-$ 4110.00 & 4120.00 $-$ 4140.00\\  
H $\gamma$ & 4270.00 $-$ 4320.00 & 4331.69 $-$ 4350.00 & 4360.00 $-$ 4410.00\\  
H $\alpha$ & 6500.00 $-$ 6550.00 & 6557.61 $-$ 6571.61 & 6575.00 $-$ 6625.00\\ 
\enddata
\end{deluxetable} 

\clearpage

\begin{deluxetable}{rrrrrrrrrrrrrr}
\tablecolumns{14} 
\tablewidth{0pc} 
\tabletypesize{\scriptsize}
\tablecaption{Equivalent Widths Measured from DIS Data}
\tablehead{\colhead{[RHG95]} & \colhead{Name} & \colhead{Ca II K} & \colhead{$\sigma_{K}$} & \colhead{H$\epsilon$} & \colhead{$\sigma_{H\epsilon}$} & \colhead{H8} & \colhead{$\sigma_{H8}$} & \colhead{H$\delta$} & \colhead{$\sigma_{H\delta}$} & \colhead{H$\gamma$} & \colhead{$\sigma_{H\gamma}$} & \colhead{H$\alpha$} & \colhead{$\sigma_{H\alpha}$} \\ 
\colhead{} & \colhead{} & \colhead{($\mbox{\AA}$)} & \colhead{($\mbox{\AA}$)} & \colhead{($\mbox{\AA}$)} & \colhead{($\mbox{\AA}$)} & \colhead{($\mbox{\AA}$)} & \colhead{($\mbox{\AA}$)} & \colhead{($\mbox{\AA}$)} & \colhead{($\mbox{\AA}$)} & \colhead{($\mbox{\AA}$)} & \colhead{($\mbox{\AA}$)} & \colhead{($\mbox{\AA}$)} & \colhead{($\mbox{\AA}$)} } 

\startdata
460 & GJ 3168 & 0.53 & 0.20 & $-$5.00 & 0.18 & $-$0.37 & 0.43 & $-$1.00 & 0.27 & 2.24 & 0.12 & $-$0.14 & 0.08 \\
492 & Gl 109 & 0.39 & 0.21 & $-$5.06 & 0.19 & $-$0.53 & 0.45 & $-$1.13 & 0.30 & 1.79 & 0.14 & 0.00 & 0.07 \\
836 & Gl 179 & 1.06 & 0.29 & $-$4.38 & 0.18 & $-$0.56 & 0.41 & $-$1.16 & 0.29 & 1.86 & 0.12 & $-$0.23 & 0.11 \\
876 & GJ 3333 & 0.46 & 0.15 & $-$5.01 & 0.15 & $-$1.48 & 0.38 & $-$1.21 & 0.28 & 2.61 & 0.11 & 0.01 & 0.07 \\
919 & Gl 203 & 0.12 & 0.29 & $-$5.30 & 0.24 & $-$0.72 & 0.41 & $-$1.84 & 0.34 & 1.98 & 0.16 & 0.09 & 0.08 \\
933 & Gl 206 & 8.55 & 0.36 & 7.55 & 0.21 & 4.12 & 0.36 & 2.97 & 0.33 & 8.78 & 0.14 & 4.57 & 0.10 \\
941 & GJ 3356 & 0.50 & 0.30 & $-$4.46 & 0.16 & $-$0.68 & 0.41 & $-$1.60 & 0.31 & 1.79 & 0.14 & $-$0.09 & 0.08 \\
1068 & GJ 3404A & 0.18 & 0.29 & $-$5.09 & 0.21 & $-$0.54 & 0.45 & $-$1.32 & 0.32 & 1.80 & 0.14 & $-$0.08 & 0.08 \\
1089 & GJ 3412 & 0.28 & 0.23 & $-$5.46 & 0.20 & $-$0.34 & 0.48 & $-$0.94 & 0.29 & 1.62 & 0.13 & $-$0.53 & 0.10 \\
1094 & Gl 251 & 0.19 & 0.24 & $-$5.01 & 0.19 & $-$0.33 & 0.44 & $-$1.28 & 0.29 & 1.97 & 0.13 & $-$0.11 & 0.08 \\
1126 & Gl 263 & 0.35 & 0.29 & $-$5.05 & 0.22 & $-$1.10 & 0.47 & $-$0.30 & 0.32 & 2.09 & 0.15 & $-$0.07 & 0.08 \\
1168 & Gl 273 & 0.37 & 0.21 & $-$5.27 & 0.17 & $-$1.40 & 0.38 & $-$1.11 & 0.30 & 2.38 & 0.12 & $-$0.09 & 0.11 \\
1184 & Gl 277B & 5.11 & 0.16 & $-$0.38 & 0.19 & 0.16 & 0.36 & 0.14 & 0.30 & 3.61 & 0.16 & 0.94 & 0.09 \\
1247 & GJ 1105 & 0.62 & 0.23 & $-$4.24 & 0.17 & $-$1.55 & 0.39 & $-$0.69 & 0.29 & 2.50 & 0.13 & $-$0.07 & 0.09 \\
1328 & GJ 2069A & 16.91 & 0.29 & 16.82 & 0.18 & 8.77 & 0.25 & 7.14 & 0.27 & 13.02 & 0.13 & 6.02 & 0.10 \\
1353 & GJ 3511B & 0.26 & 0.30 & $-$5.68 & 0.28 & $-$1.06 & 0.56 & $-$0.11 & 0.30 & 2.57 & 0.16 & $-$0.12 & 0.06 \\
1405 & GJ 3522 & 7.10 & 0.21 & 4.99 & 0.17 & 1.90 & 0.30 & 1.53 & 0.33 & 7.08 & 0.11 & 2.76 & 0.10 \\
1420 & GJ 3527 & 0.15 & 0.25 & $-$5.91 & 0.17 & $-$1.78 & 0.38 & $-$1.64 & 0.33 & 2.22 & 0.16 & $-$0.22 & 0.09 \\
1441 & GJ 3537 & 7.28 & 0.15 & 5.73 & 0.30 & 3.70 & 0.53 & 3.88 & 0.28 & 7.92 & 0.21 & 3.79 & 0.08 \\
1501 & GJ 1125 & 0.25 & 0.28 & $-$4.79 & 0.17 & $-$1.31 & 0.40 & $-$1.61 & 0.32 & 2.14 & 0.14 & 0.00 & 0.08 \\
1507 & Gl 352A & 0.19 & 0.19 & $-$5.87 & 0.22 & $-$1.37 & 0.40 & $-$0.39 & 0.32 & 2.00 & 0.17 & 0.00 & 0.07 \\
1529 & Gl 362 & 3.73 & 0.29 & 1.75 & 0.20 & 3.43 & 0.30 & 3.05 & 0.26 & 6.19 & 0.14 & 1.10 & 0.10 \\
1530 & Gl 363 & 0.52 & 0.21 & $-$4.28 & 0.19 & $-$0.92 & 0.39 & $-$0.48 & 0.27 & 2.40 & 0.13 & $-$0.19 & 0.10 \\
1611 & Gl 386 & 0.26 & 0.33 & $-$6.07 & 0.24 & $-$0.88 & 0.43 & $-$0.09 & 0.32 & 2.25 & 0.18 & $-$0.10 & 0.07 \\
1616 & Gl 388 & 9.64 & 0.27 & 6.54 & 0.22 & 3.20 & 0.33 & 3.02 & 0.27 & 6.82 & 0.16 & 3.73 & 0.08 \\
1657 & GJ 3612 & 0.03 & 0.27 & $-$6.20 & 0.24 & $-$0.88 & 0.44 & $-$0.51 & 0.31 & 2.38 & 0.16 & 0.13 & 0.09 \\
1658 & Gl 398 & 10.40 & 0.25 & 9.03 & 0.23 & 5.79 & 0.34 & 6.10 & 0.31 & 10.66 & 0.17 & 5.14 & 0.09 \\
1660 & GJ 3613 & 9.86 & 0.25 & 7.09 & 0.25 & 3.76 & 0.38 & 3.41 & 0.27 & 7.38 & 0.19 & 3.96 & 0.09 \\
1690 & Gl 403 & 2.85 & 0.36 & $-$5.15 & 0.24 & $-$1.27 & 0.46 & $-$0.49 & 0.33 & 2.45 & 0.16 & $-$0.20 & 0.08 \\
1742 & GJ 3647 & 9.41 & 0.32 & 9.25 & 0.27 & 7.10 & 0.38 & 6.43 & 0.36 & 11.08 & 0.19 & 5.39 & 0.08 \\
1795 & GJ 3666 & 0.21 & 0.24 & $-$5.72 & 0.24 & $-$1.29 & 0.40 & $-$0.39 & 0.32 & 2.24 & 0.11 & 0.05 & 0.09 \\
1838 & Gl 443 & 0.92 & 0.15 & $-$4.89 & 0.24 & 0.07 & 0.45 & $-$0.39 & 0.25 & 2.23 & 0.17 & $-$0.22 & 0.08 \\
1845 & Gl 445 & 0.05 & 0.31 & $-$6.10 & 0.24 & $-$1.27 & 0.42 & $-$0.46 & 0.31 & 2.25 & 0.13 & $-$0.73 & 0.13 \\
1931 & GJ 3719 & 9.95 & 0.21 & 8.29 & 0.25 & 6.88 & 0.47 & 5.37 & 0.26 & 10.16 & 0.16 & 4.83 & 0.09 \\
1988 & Gl 480 & 0.59 & 0.20 & $-$5.28 & 0.22 & $-$1.37 & 0.37 & $-$0.34 & 0.32 & 2.05 & 0.16 & 0.05 & 0.08 \\
2009 & Gl 486 & 0.17 & 0.24 & $-$5.68 & 0.32 & $-$1.68 & 0.58 & $-$0.36 & 0.33 & 1.75 & 0.16 & 0.01 & 0.09 \\
2088 & GJ 3775 & 0.61 & 0.31 & $-$4.13 & 0.17 & $-$1.53 & 0.35 & $-$0.12 & 0.31 & 2.29 & 0.15 & $-$0.13 & 0.10 \\
2094 & Gl 507B & 0.26 & 0.33 & $-$4.41 & 0.21 & $-$1.56 & 0.39 & $-$0.51 & 0.31 & 2.40 & 0.18 & $-$0.13 & 0.09 \\
2175 & GJ 3804 & 0.31 & 0.38 & $-$4.42 & 0.21 & $-$1.19 & 0.40 & $-$0.34 & 0.30 & 2.57 & 0.20 & $-$0.12 & 0.10 \\
2300 & Gl 553.1 & 0.14 & 0.28 & $-$4.79 & 0.19 & $-$1.27 & 0.43 & $-$0.56 & 0.29 & 2.53 & 0.15 & $-$0.20 & 0.10 \\
2310 & Gl 555 & 0.77 & 0.26 & $-$3.41 & 0.17 & $-$1.42 & 0.38 & $-$0.39 & 0.28 & 2.31 & 0.14 & $-$0.17 & 0.11 \\
2399 & GJ 3892 & 0.40 & 0.29 & $-$4.93 & 0.20 & $-$1.37 & 0.34 & $-$0.43 & 0.30 & 2.41 & 0.18 & $-$0.24 & 0.09 \\
2410 & Gl 581 & 0.08 & 0.32 & $-$4.82 & 0.17 & $-$1.58 & 0.38 & $-$0.50 & 0.31 & 2.36 & 0.18 & $-$0.17 & 0.08 \\
2456 & GJ 3913 & 0.17 & 0.32 & $-$5.03 & 0.19 & $-$1.73 & 0.36 & $-$0.54 & 0.30 & 2.66 & 0.15 & $-$0.45 & 0.12 \\
2472 & Gl 597 & 0.20 & 0.37 & $-$4.81 & 0.18 & $-$1.56 & 0.42 & $-$0.74 & 0.32 & 2.43 & 0.25 & $-$0.13 & 0.09 \\
2567 & Gl 617B & 0.63 & 0.28 & $-$4.58 & 0.19 & $-$1.35 & 0.36 & $-$0.53 & 0.31 & 2.31 & 0.19 & $-$0.27 & 0.08 \\
2587 & GJ 3953 & 1.16 & 0.23 & $-$4.00 & 0.18 & 0.06 & 0.40 & $-$0.52 & 0.31 & 2.48 & 0.15 & 0.10 & 0.09 \\
2599 & Gl 628 & 0.71 & 0.28 & $-$4.86 & 0.17 & $-$0.77 & 0.41 & $-$0.42 & 0.28 & 2.20 & 0.10 & $-$0.27 & 0.10 \\
2660 & Gl 643 & 3.79 & 0.26 & 0.20 & 0.20 & 2.48 & 0.33 & 2.11 & 0.28 & 5.58 & 0.12 & 2.21 & 0.07 \\
2661 & Gl 644A & 0.44 & 0.51 & $-$4.34 & 0.26 & $-$0.58 & 0.45 & $-$0.69 & 0.37 & 2.33 & 0.14 & $-$0.14 & 0.10 \\
2665 & GJ 1207 & 7.49 & 0.39 & 6.11 & 0.33 & 4.23 & 0.49 & 2.61 & 0.35 & 8.05 & 0.19 & 3.94 & 0.10 \\
2700 & Gl 655 & 0.45 & 0.18 & $-$5.10 & 0.18 & $-$0.71 & 0.39 & $-$0.45 & 0.28 & 2.11 & 0.13 & $-$0.37 & 0.08 \\
2712 & Gl 660A & 0.56 & 0.24 & $-$5.22 & 0.22 & $-$1.29 & 0.44 & $-$0.78 & 0.36 & 2.22 & 0.16 & 0.08 & 0.09 \\
2714 & GJ 3992 & 0.47 & 0.17 & $-$5.06 & 0.18 & $-$0.92 & 0.37 & $-$0.45 & 0.27 & 2.04 & 0.12 & $-$0.41 & 0.10 \\
2717 & Gl 661A & 0.06 & 0.28 & $-$5.88 & 0.21 & $-$0.97 & 0.41 & $-$0.60 & 0.29 & 2.50 & 0.16 & $-$0.34 & 0.09 \\
2719 & GJ 1212 & 0.22 & 0.32 & $-$4.13 & 0.18 & $-$1.55 & 0.39 & $-$0.31 & 0.29 & 2.45 & 0.16 & $-$0.20 & 0.09 \\
2744 & Gl 669A & 4.45 & 0.25 & 0.58 & 0.17 & 1.36 & 0.30 & 0.94 & 0.31 & 4.42 & 0.14 & 1.48 & 0.12 \\
2797 & Gl 687 & 0.14 & 0.14 & $-$4.77 & 0.27 & $-$1.35 & 0.39 & $-$0.76 & 0.41 & 3.00 & 0.15 & $-$0.23 & 0.11 \\
2846 & GJ 4038 & 0.42 & 0.18 & $-$5.33 & 0.18 & $-$0.67 & 0.41 & $-$0.55 & 0.28 & 1.97 & 0.12 & $-$0.16 & 0.08 \\
2849 & Gl 699 & 0.23 & 0.46 & $-$4.42 & 0.23 & $-$1.52 & 0.44 & $-$0.86 & 0.34 & 2.93 & 0.13 & $-$0.78 & 0.13 \\
2851 & GJ 4040 & 1.13 & 0.24 & $-$4.19 & 0.20 & $-$0.17 & 0.43 & $-$0.16 & 0.33 & 2.30 & 0.15 & 0.15 & 0.08 \\
2888 & GJ 4048 & 0.12 & 0.22 & $-$5.88 & 0.30 & $-$1.45 & 0.53 & $-$1.07 & 0.34 & 1.56 & 0.18 & $-$0.21 & 0.09 \\
2895 & GJ 1226 & 0.13 & 0.20 & $-$6.07 & 0.25 & $-$0.68 & 0.47 & $-$1.14 & 0.37 & 1.76 & 0.16 & $-$0.09 & 0.09 \\
2898 & Gl 712 & 0.09 & 0.34 & $-$5.53 & 0.22 & $-$1.34 & 0.38 & $-$0.69 & 0.31 & 2.76 & 0.13 & $-$0.19 & 0.10 \\
2901 & GJ 4055 & 0.15 & 0.23 & $-$5.42 & 0.23 & $-$0.89 & 0.45 & $-$0.79 & 0.33 & 1.97 & 0.15 & $-$0.09 & 0.10 \\
2916 & GJ 4062 & 0.66 & 0.22 & $-$4.72 & 0.22 & $-$1.22 & 0.46 & $-$0.45 & 0.34 & 1.77 & 0.14 & $-$0.13 & 0.09 \\
2921 & GJ 4063 & 1.10 & 0.28 & $-$4.24 & 0.18 & $-$0.96 & 0.39 & $-$0.33 & 0.28 & 2.20 & 0.13 & 0.08 & 0.09 \\
2924 & Gl 720B & 0.29 & 0.26 & $-$5.04 & 0.20 & $-$1.28 & 0.40 & $-$0.43 & 0.30 & 2.10 & 0.12 & $-$0.13 & 0.10 \\
2937 & GJ 4068 & 5.76 & 0.39 & 3.74 & 0.24 & 0.98 & 0.33 & 2.21 & 0.33 & 6.45 & 0.17 & 2.82 & 0.09 \\
2940 & GJ 4070 & 0.21 & 0.19 & $-$5.50 & 0.19 & $-$0.84 & 0.40 & $-$0.43 & 0.28 & 2.18 & 0.12 & $-$0.14 & 0.07 \\
2945 & Gl 725A & $-$0.12 & 0.45 & $-$5.35 & 0.32 & $-$0.79 & 0.56 & $-$0.87 & 0.43 & 2.80 & 0.29 & $-$0.29 & 0.11 \\
2946 & Gl 725B & 0.10 & 0.26 & $-$5.98 & 0.21 & $-$0.28 & 0.47 & $-$0.58 & 0.32 & 2.28 & 0.13 & $-$0.12 & 0.07 \\
2964 & GJ 4083 & 0.16 & 0.24 & $-$6.34 & 0.47 & $-$0.84 & 0.45 & $-$0.59 & 0.29 & 2.25 & 0.14 & $-$0.19 & 0.09 \\
2976 & Gl 735 & 5.76 & 0.26 & 1.52 & 0.17 & 1.67 & 0.27 & 1.07 & 0.30 & 4.59 & 0.14 & 2.29 & 0.08 \\
3013 & GJ 4098 & 0.22 & 0.31 & $-$4.89 & 0.18 & $-$1.31 & 0.40 & $-$0.44 & 0.30 & 2.49 & 0.16 & $-$0.49 & 0.10 \\
3032 & GJ 9652A & 10.77 & 0.27 & 9.09 & 0.22 & 4.26 & 0.30 & 4.19 & 0.29 & 9.71 & 0.18 & 5.40 & 0.11 \\
3058 & GJ 4110 & 5.84 & 0.28 & 2.29 & 0.20 & 2.06 & 0.29 & 2.29 & 0.32 & 6.35 & 0.14 & 2.74 & 0.11 \\
3306 & GJ 9721B & 3.73 & 0.59 & 3.26 & 0.37 & $-$0.17 & 1.05 & 2.33 & 0.39 & 6.38 & 0.31 & 2.69 & 0.09 \\
3425 & GJ 4231 & 11.03 & 0.29 & 12.54 & 0.20 & 4.03 & 0.31 & 3.77 & 0.27 & 6.84 & 0.18 & 4.18 & 0.08 \\
3478 & Gl 849 & 0.72 & 0.21 & $-$5.37 & 0.23 & $-$0.46 & 0.44 & $-$0.14 & 0.30 & 2.03 & 0.17 & $-$0.37 & 0.10 \\
3584 & Gl 873 & 8.95 & 0.30 & 6.26 & 0.22 & 4.99 & 0.34 & 4.99 & 0.33 & 9.02 & 0.16 & 3.97 & 0.10 \\
\enddata
\end{deluxetable}

\clearpage
\begin{deluxetable}{rrrrrrrrrr}
\tablecolumns{10}  
\tablewidth{0pc}  
\tablecaption{Equivalent Widths Measured from Echelle Data}
\tabletypesize{\scriptsize}
\tablehead{\colhead{[RHG95]} & \colhead{Name} & \colhead{Ca II K$_{ARCES}$} & \colhead{$\sigma_{K, ARCES}$} & \colhead{H$\alpha_{ARCES}$} & \colhead{$\sigma_{H\alpha, ARCES}$} & \colhead{Ca II K$_{HIRES}$} & \colhead{$\sigma_{K, HIRES}$} & \colhead{$H\alpha_{HIRES}$} & \colhead{$\sigma_{H\alpha, HIRES}$} \\ 
\colhead{} & \colhead{} & \colhead{($\mbox{\AA}$)} & \colhead{($\mbox{\AA}$)} & \colhead{($\mbox{\AA}$)} & \colhead{($\mbox{\AA}$)} & \colhead{($\mbox{\AA}$)} & \colhead{($\mbox{\AA}$)} & \colhead{($\mbox{\AA}$)} & \colhead{($\mbox{\AA}$)} } 
\startdata
492 & Gl 109 & \nodata & \nodata & \nodata & \nodata & 0.35 & 0.03 & $-$0.24 & 0.00 \\
836 & Gl 179 & \nodata & \nodata & \nodata & \nodata & 0.72 & 0.07 & $-$0.14 & 0.00 \\
1089 & GJ 3412 & 0.20 & 0.04 & $-$0.28 & 0.03 & \nodata & 0.00 & \nodata & \nodata \\
1094 & Gl 251 & 0.23 & 0.01 & $-$0.27 & 0.03 & 0.22 & 0.03 & $-$0.23 & 0.03 \\
1126 & Gl 263 & 0.34 & 0.02 & $-$0.30 & 0.02 & \nodata & \nodata & \nodata & \nodata \\
1168 & Gl 273 & \nodata & \nodata & \nodata & \nodata & 0.26 & 0.04 & $-$0.17 & 0.02 \\
1328 & GJ 2069A & 2.25 & 0.26 & 2.51 & 0.03 & \nodata & \nodata & \nodata & \nodata \\
1501 & GJ 1125 & \nodata & \nodata & \nodata & \nodata & 0.17 & 0.04 & $-$0.19 & 0.02 \\
1529 & Gl 362 & 1.65 & 0.10 & 0.53 & 0.03 & 1.68 & 0.28 & 0.54 & 0.01 \\
1616 & Gl 388 & \nodata & \nodata & \nodata & \nodata & 4.93 & 1.79 & 2.26 & 0.76 \\
1658 & Gl 398 & 2.71 & 0.19 & 3.23 & 0.03 & \nodata & \nodata & \nodata & \nodata \\
1690 & Gl 403 & 0.15 & 0.02 & $-$0.23 & 0.03 & \nodata & \nodata & \nodata & \nodata \\
1838 & Gl 443 & 0.46 & 0.08 & $-$0.30 & 0.03 & \nodata & \nodata & \nodata & \nodata \\
1845 & Gl 445 & \nodata & \nodata & \nodata & \nodata & 0.12 & 0.01 & $-$0.24 & 0.00 \\
1988 & Gl 480 & 0.57 & 0.06 & $-$0.34 & 0.03 & \nodata & \nodata & \nodata & \nodata \\
2009 & Gl 486 & \nodata & \nodata & \nodata & \nodata & 0.17 & 0.02 & $-$0.16 & 0.00 \\
2175 & GJ 3804 & \nodata & \nodata & \nodata & \nodata & 0.27 & 0.03 & $-$0.14 & 0.00 \\
2310 & Gl 555 & 0.20 & 0.02 & $-$0.18 & 0.02 & \nodata & \nodata & \nodata & \nodata \\
2410 & Gl 581 & 0.09 & 0.01 & $-$0.22 & 0.02 & 0.19 & 0.05 & $-$0.21 & 0.04 \\
2567 & Gl 617B & \nodata & \nodata & \nodata & \nodata & 0.58 & 0.05 & $-$0.34 & 0.00 \\
2587 & GJ 3953 & 0.67 & 0.07 & $-$0.10 & 0.02 & \nodata & \nodata & \nodata & \nodata \\
2599 & Gl 628 & 0.18 & 0.01 & $-$0.21 & 0.02 & 0.27 & 0.03 & $-$0.21 & 0.01 \\
2660 & Gl 643 & 0.15 & 0.03 & $-$0.19 & 0.02 & \nodata & \nodata & \nodata & \nodata \\
2665 & GJ 1207 & 2.38 & 0.18 & 2.56 & 0.04 & \nodata & \nodata & \nodata & \nodata \\
2700 & Gl 655 & \nodata & \nodata & \nodata & \nodata & 0.28 & 0.03 & $-$0.27 & 0.00 \\
2712 & Gl 660A & 0.38 & 0.03 & $-$0.27 & 0.03 & \nodata & \nodata & \nodata & \nodata \\
2714 & GJ 3992 & \nodata & \nodata & \nodata & \nodata & 0.38 & 0.05 & $-$0.26 & 0.01 \\
2797 & Gl 687 & \nodata & \nodata & \nodata & \nodata & 0.31 & 0.07 & $-$0.27 & 0.02 \\
2851 & GJ 4040 & 0.48 & 0.05 & $-$0.21 & 0.02 & \nodata & \nodata & \nodata & \nodata \\
2921 & GJ 4063 & \nodata & \nodata & \nodata & \nodata & 0.61 & 0.07 & $-$0.19 & 0.00 \\
2940 & GJ 4070 & \nodata & \nodata & \nodata & \nodata & 0.14 & 0.01 & $-$0.24 & 0.00 \\
2945 & Gl 725A & \nodata & \nodata & \nodata & \nodata & 0.15 & 0.04 & $-$0.23 & 0.00 \\
2946 & Gl 725B & \nodata & \nodata & \nodata & \nodata & 0.19 & 0.03 & $-$0.24 & 0.02 \\
3013 & GJ 4098 & \nodata & \nodata & \nodata & \nodata & 0.18 & 0.02 & $-$0.24 & 0.00 \\
3478 & Gl 849 & \nodata & \nodata & \nodata & \nodata & 0.42 & 0.19 & $-$0.34 & 0.02 \\
\enddata
\end{deluxetable}
\clearpage

\begin{deluxetable}{rrrrrr}
\tablecolumns{6}  
\tablewidth{0pc}  
\tablecaption{Stars with UV and X-ray Measurements}
\tabletypesize{\footnotesize}
\tablehead{\colhead{[RHG95]} & \colhead{Name} & \colhead{F$_{ROSAT}$} & \colhead{F$_{MD89 Xray}$} & \colhead{F$_{MgII, MD89}$} & \colhead{F$_{MgII, ACS}$} \\ 
\colhead{} & \colhead{} & \multicolumn{4}{c}{LOG (ergs s$^{-1}$ cm$^{-2}$  $\mbox{\AA}^{-1}$)}} 
\startdata
492 & Gl 109 & $-$12.53 & \nodata & \nodata & \nodata \\
876 & GJ 3333 & $-$12.54 & \nodata & \nodata & \nodata \\
933 & Gl 206 & $-$11.39 & $-$11.43 & \nodata & $-$12.56 \\
941 & GJ 3356 & $-$12.74 & \nodata & \nodata & \nodata \\
1168 & Gl 273 & $-$12.99 & \nodata & $-$12.52 & $-$12.68 \\
1184 & Gl 277B & $-$11.14 & \nodata & \nodata & \nodata \\
1328 & GJ 2069A & $-$11.26 & \nodata & \nodata & \nodata \\
1405 & GJ 3522 & $-$10.93 & \nodata & \nodata & \nodata \\
1507 & Gl 352A & $-$12.97 & \nodata & \nodata & \nodata \\
1529 & Gl 362 & $-$11.96 & \nodata & \nodata & \nodata \\
1616 & Gl 388 & $-$10.60 & $-$10.50 & $-$11.90 & $-$11.72 \\
1658 & Gl 398 & $-$11.85 & \nodata & \nodata & \nodata \\
1742 & GJ 3647 & $-$11.58 & \nodata & \nodata & \nodata \\
2567 & Gl 617B & $-$12.47 & \nodata & \nodata & \nodata \\
2599 & Gl 628 & $-$12.82 & \nodata & \nodata & \nodata \\
2660 & Gl 643 & $-$10.56 & \nodata & \nodata & \nodata \\
2661 & Gl 644A & $-$10.56 & $-$11.13 & \nodata & $-$12.08 \\
2665 & GJ 1207 & $-$11.66 & \nodata & \nodata & \nodata \\
2717 & Gl 661A & $-$12.56 & \nodata & \nodata & \nodata \\
2744 & Gl 669A & $-$11.35 & \nodata & \nodata & \nodata \\
2797 & Gl 687 & $-$12.51 & \nodata & \nodata & \nodata \\
2916 & GJ 4062 & $-$13.32 & \nodata & \nodata & \nodata \\
2945 & Gl 725A & $-$12.70 & \nodata & \nodata & \nodata \\
2946 & Gl 725B & $-$12.70 & \nodata & \nodata & \nodata \\
2976 & Gl 735 & $-$11.00 & $-$11.59 & $-$12.17 & $-$12.24 \\
3478 & Gl 849 & \nodata & $-$12.67 & \nodata & $-$13.03 \\
3584 & Gl 873 & $-$10.40 & \nodata & \nodata & $-$11.92 \\
\enddata
\end{deluxetable}

\end{document}